\pdfoutput=1  %this is to force ArXiv Tex To use PDFTeXify command
\documentclass{gHPR2e}
\usepackage{subfigure}% Support for small, `sub' figures and tables

\begin{document}

\title{A low-background piston-cylinder type hybrid high pressure cell for muon-spin rotation/relaxation experiments}

\author{Z. Shermadini,$^{\rm a,}$$^{\ast}$\thanks{$^\ast$Corresponding author. Email: zurab.shermadini@psi.ch; Laboratory for Muon Spin Spectroscopy, Paul Scherrer Institut,
CH-5232 Villigen PSI, Switzerland
\vspace{6pt}}
R. Khasanov,$^{\rm a}$
M.~Elender,$^{\rm a}$
G.~Simutis,$^{\rm a}$
Z.~Guguchia,$^{\rm a,}$$^{\rm b}$
K.V.~Kamenev,$^{\rm c}$
and A.~Amato$^{\rm a}$
{\small \\\vspace{6pt}  $^{a}${\em{Laboratory for Muon Spin Spectroscopy, Paul Scherrer Institut, Villigen PSI, Switzerland}}
\\\vspace{3pt}  $^{b}${\em{Department of Physics, Columbia University, New York, NY, USA}}
\\\vspace{3pt}  $^{c}${\em{Centre for Science at Extreme Conditions and School of Engineering, University of Edinburgh, Edinburgh, United Kingdom}}}
\\\received{September 2017} }

\maketitle

%%%%%%%%%%%%%%%%%%%%%%%%%%%%%%%%%%%%%%%%%%%%%%%%End HPST

\noindent {\it Abstract:} {\small A low background double-wall piston-cylinder-type pressure cell is
developed at the Paul Scherrer Institute. The cell is made from
BERYLCO-25 (beryllium copper) and MP35N nonmagnetic alloys
with the design and dimensions which are specifically adapted to
muon-spin rotation/relaxation ($\mu$SR) measurements. The mechanical
design and performance of the pressure cell are evaluated using
finite-element analysis (FEA). By including the measured stress-strain
characteristics of the materials into the finite-element model,
the cell dimensions are optimized with the aim to reach the
highest possible pressure while maintaining the sample space
large (6~mm in diameter and 12~mm high). The presented
unconventional design of the double-wall piston-cylinder pressure
cell with a harder outer MP35N sleeve and a softer inner CuBe
cylinder enables pressures of up to 2.6~GPa to be reached at
ambient temperature, corresponding to 2.2 GPa at low
temperatures without any irreversible damage to the pressure cell.
The nature of the muon stopping distribution, mainly in the sample
and in the CuBe cylinder, results in a low-background $\mu$SR signal.}

\vspace{0.2cm}

\section{Introduction}
A hydrostatic piston-cylinder type pressure cell is favorable for $\mu$SR measurements due to a large sample volume and the compactness of the design \cite{Khasanov1}. It consists, typically, of a cylindrically shaped body, mushroom type sealing system, compressing pistons, and locking nut(s). Using an external force, the piston compresses a pressure transmitting liquid and generates high pressure on the sample. The highest pressure achieved by this system depends on the sample space dimensions and the materials used in the body-cell cylinders. Close to 5~GPa pressure is reported in such a type of cell using CuBe and NiCrAl alloys with a sample space diameter of 4.4~mm and outer sleeve diameter of 40~mm~\cite{Fujiwara,Taniguchi}. For inelastic neutron scattering experiments, a pressure up to 1.8~GPa is reached at room temperature in a double-wall cell with 6~mm inner bore and 18~mm outer sleeve diameter~\cite{Kamenev1}. The piston type cells are for some techniques essential giving the possibility to reduce the background signal.

For a successful $\mu$SR experiment, a low background signal is required. Using high-energy muons, it is possible to maximize a sample size while minimizing a pressure cell wall thickness to reduce a scattering of the implanted muons and hence maintain a good sample/background signal ratio. In addition, the materials used for cell production have to be nonmagnetic with minimal temperature and field dependent relaxation rates. The most favorable candidates are CuBe and MP35N alloys. With our previously reported double-wall pressure cell made completely from MP35N material, one can reach 2.4~GPa at low temperatures (for more details see~\cite{Khasanov1}).
Despite high reachable pressures in such a type of cell, MP35N alloy has an increased relaxation rate below 1~K. Bearing in mind that at least half of all the implanted muons stop in the pressure cell walls, this complicates the analysis of $\mu$SR data in the low-temperature region. Therefore, a material with low relaxation rate such as CuBe is favorable.

\begin{figure}[t]
\centering
\includegraphics[width=0.34\linewidth]{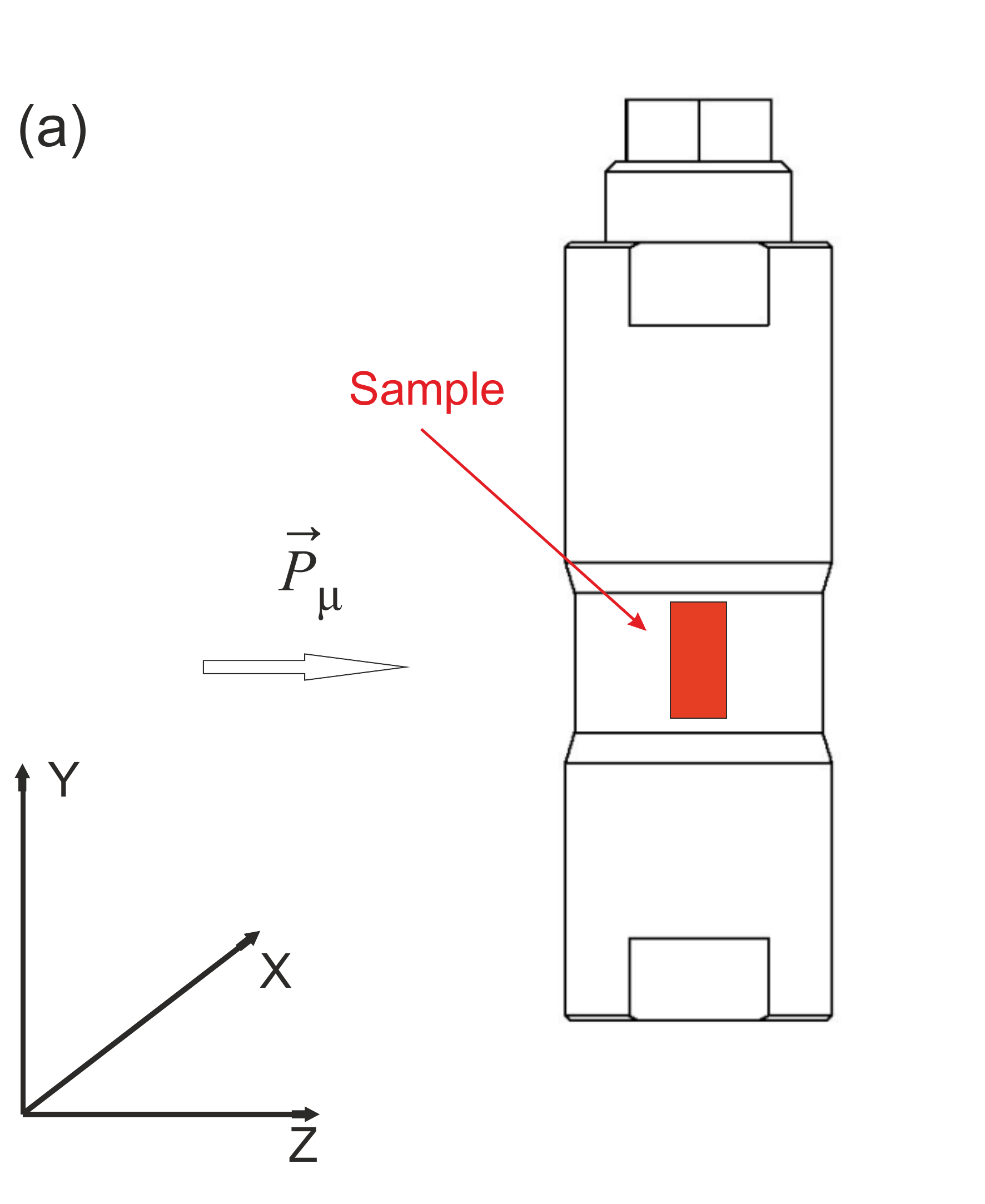}~~~~
\includegraphics[width=0.47\linewidth]{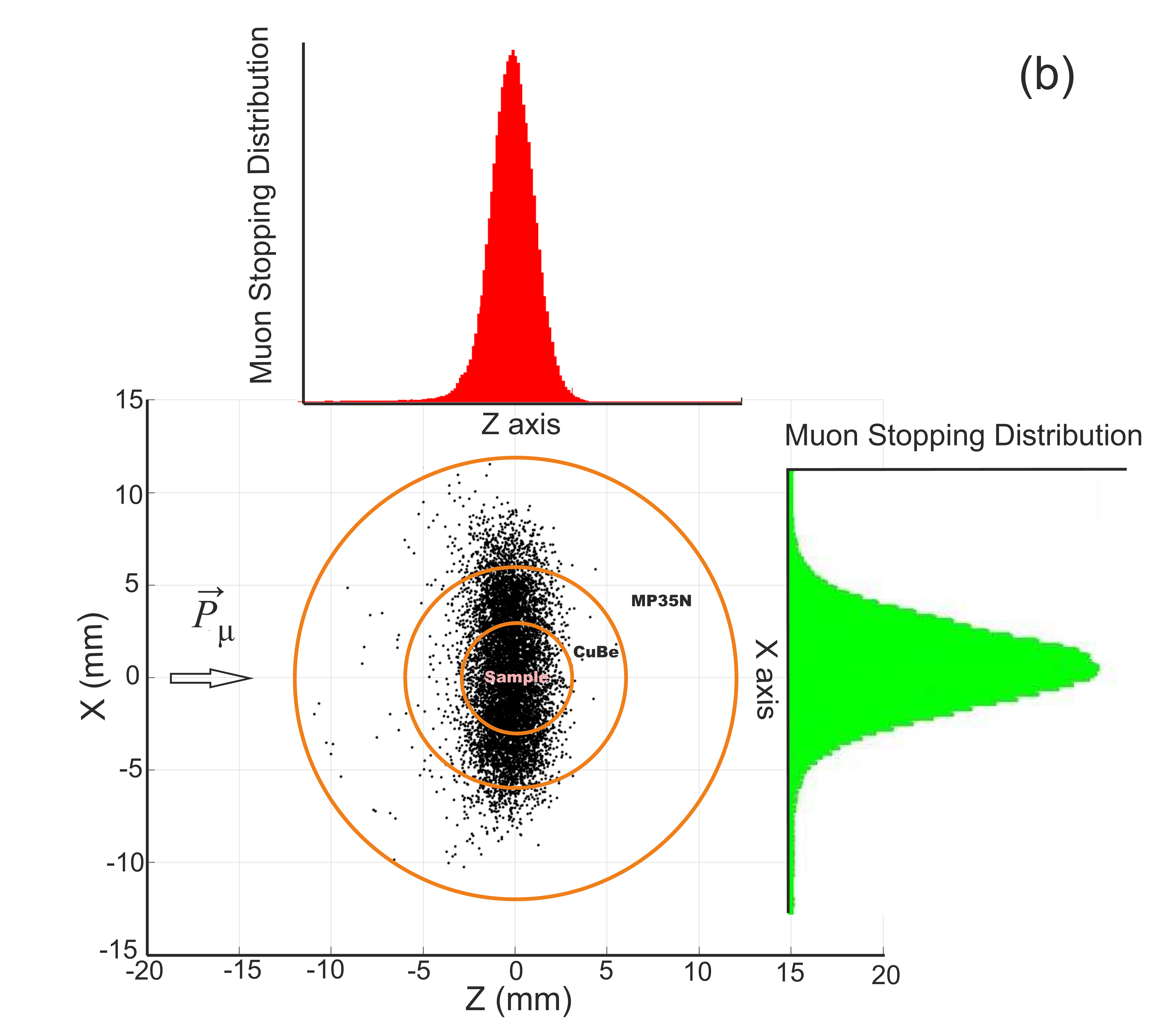}
%\vspace{-7.5cm}
\caption{\footnotesize  (a) The schematic view of cylindrical pressure cell (black contour) with the sample inside (red rectangular). Muons are implanted along the vector $P_{\mu}$. (b) The cross sectional view ($X-Z$ plane) of the pressure cell consisting of the inner and the outer cylinders made of CuBe and MP35N alloys. The colored areas represent the muon stopping distributions in parallel (red) and perpendicular (green) directions to the muon beam. The energy of implanted muons was 44~MeV. The simulations were performed by using TRIM.SP package, Ref.~\cite{SRIM}. The simulations revealed that approximately 37\% of all the muons stop within the sample, $\simeq 43$\% within the inner and $\simeq 10$\% within the outer cylinder, respectively.}
\label{Fig:SRIM}
\end{figure}

It should be noted, however, that the ultimate tensile strength of the CuBe alloy is approximately half of that for the MP35N, which prevents reaching high pressures by using cells made purely of CuBe material. There is, however, a possibility to overcome such difficulty based on the important feature of the muon-spin rotation/relaxation experiment. The point is that by penetrating the sample, the muon gradually loses energy through multiple interactions and eventually stops at a certain  distance from the sample surface.  By stopping, the muon further decays by emitting the positron(electron). Consequently the information obtained in $\mu$SR experiment becomes 'depth selected`. This is of fundamental importance to the low-energy muon experiments Ref.~\cite{LEM1, LEM2}, where muons with tunable energy (from 0 up to $\simeq 30$~keV) are allowed to stop at a controllable distance from the sample surface (from 0 up to $\sim 200$~nm). The stopping profile of muons is determined by the energy of the muons in the beam. In high pressure experiments, much higher energies of muons are needed (10-100 MeV) to penetrate the pressure cell and be stopped in the sample. A schematic view of $\mu$SR under pressure experiments and the distribution of muons stopped within the pressure cell and the sample are presented in Figure~\ref{Fig:SRIM}. The calculations were performed by using TRIM.SP package, Ref.~~\cite{SRIM}, which is widely used to determine the stopping profile of charge particles implanted into the solid. The real dimensions of the cell were used as input parameters for the simulation with the inner CuBe cylinder (inner/outer diameter of 6/12 mm) and outer MP35N (12/24 mm) cylinder. The sample (6 mm in diameter and 12 mm in height) was assumed to be made out of CuBe. The implanted muon energy was set to 44~MeV. The muon beam cross section  (4 x 10 mm$^2$) and the cryostat walls, which muons need to pass before reaching the cell, correspond to the experimental setup at the GPD (General Purpose Decay) instrument at the Paul Scherrer Institute (PSI, Villigen, Switzerland). By properly selecting the muon implantation energy, most of muons could be stopped within the sample and the inner cylinder area, while only the minor fraction of all the muons would hit the outer cylinder of the cell (see Figure~\ref{Fig:SRIM}~(b)).
This implies that the double-wall cell where only the inner cylinder is made out of CuBe alloy could be still considered as the low-background pressure cell for $\mu$SR experiment.

In this paper, we describe a modified design of a hybrid low background piston-cylinder type pressure cell for muon-spin rotation/relaxation experiments. The double-wall piston-cylinder pressure cell with an outer MP35N sleeve and inner CuBe cylinder allows us to safely reach pressures up to 2.2 GPa at low temperatures. The muon stopping distribution character, mainly in the sample and in the inner CuBe cylinder, leads to $\mu$SR signal with reduced background. The first generation PSI pressure cells were single-wall CuBe cylinder clamp cells~\cite{Andreica}. This cell with inner sample space diameter of 5~mm could reach the pressures up to 1.4~GPa at low temperatures. The advantage of the new double-wall CuBe/MP35N
cell compared to the previous single-wall CuBe cylinder clamp cell is pressure performance
and increased safety factor due to MP35N outer sleeve having higher strength and lower
brittleness compared to BERYLCO-25. Using softer material in the construction of the
inner cylinder is a novel and unconventional approach to the design of double-layered
piston-cylinder cells. It works by allowing large plastic strain in the inner cylinder while constraining
its deformation with a harder outer cylinder.

The paper is organized as follows.  The finite-element analysis (FEA) applied to optimization of cell dimensions in order to reach a maximum pressure is described in Section.~\ref{sec:FEA}. The basic approach of the simulation and optimization methods is shown, and comparisons with real test measurements are presented. In Section.~\ref{sec:design_and_assembly}, the cell design and assembly details are discussed. The test of the pressure is described in Section.~\ref{sec:cell-test}. The results of $\mu$SR studies of the pressure cell and Si$_3$N$_4$ pistons are summarised in Section.~\ref{sec:background-cntributions}. Measurements of the pressure induced superconductivity in CrAs compound and their comparison with the previously published data are given in Section.~\ref{sec:example}. The conclusions follow in Section.~\ref{sec:conclusions}.

\section{Finite-element analysis}\label{sec:FEA}

\begin{figure}[h]
	\centering
	\includegraphics[width=0.49\linewidth]{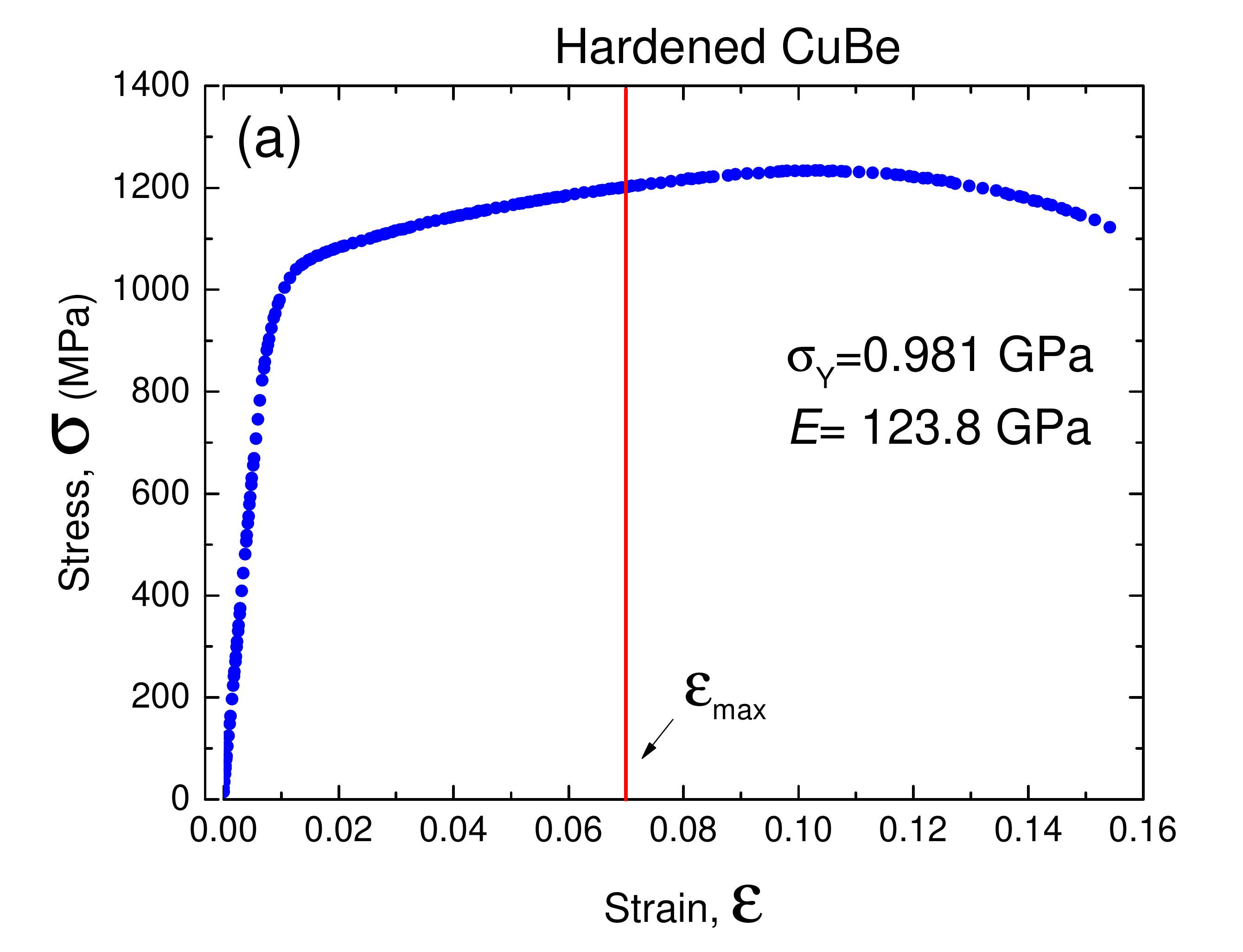}~~~~
	\includegraphics[width=0.49\linewidth]{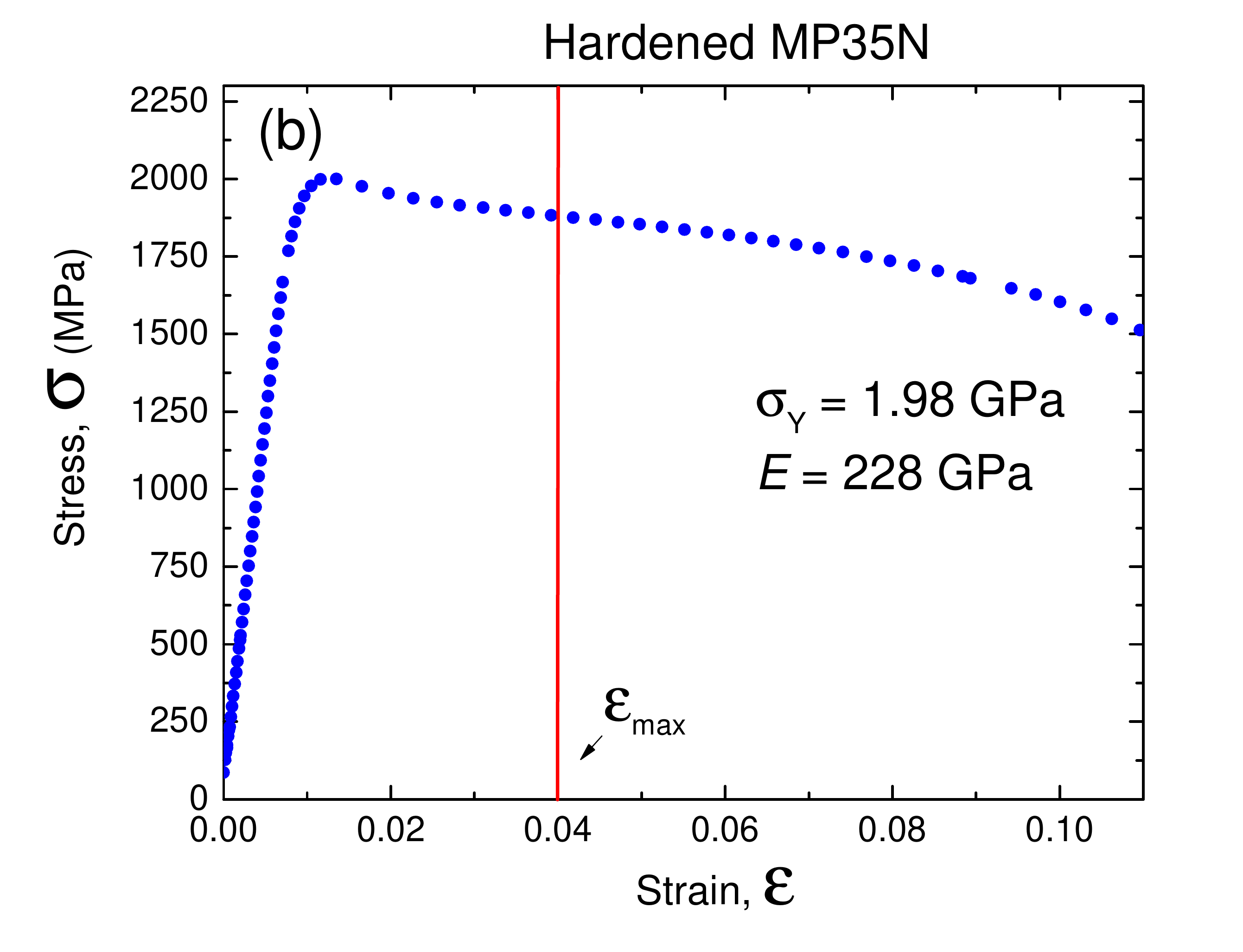}
	%\vspace{-7.5cm}
	\caption{\footnotesize Engineering stress-strain measurement curves: (a) Hardened CuBe; (b) Hardened MP35N. Red lines represent maximum strains allowed to be developed in CuBe ($\varepsilon_{\rm CuBe} \leq $ 0.07 mm/mm) and MP35N ( $\varepsilon_{\rm MP35N} \leq $ 0.04 mm/mm) cylinders during FEA optimization.  }
	\label{Fig:StressStrain}
\end{figure}

The finite-element analysis was applied to simulate the response of the cell on applied hydrostatic pressure. Optimizations of the design were performed within the so-called non-linear regime, which is based on the maximum yield strain criteria. The corresponding engineering stress-strain curves for the very same CuBe and MP35N alloys used for the double-wall cell production were specifically measured at PSI (see Figure~\ref{Fig:StressStrain}).
 The measured engineering stress-strain relations were converted into the true stress-strain data using the relations $\sigma_{t}=\sigma_{e}(1+\varepsilon_{e})$, $\varepsilon_{t}=\ln(1+\varepsilon_{e})$, and then the extracted plastic deformation parts were inserted to ANSYS Workbench~\cite{ANSYS} as the multilinear isotropic hardening material properties. The bulk modulus was extracted from the linear parts of the true stress-strain curves. The remaining parameters such as Poisson's ratio and shear modulus were taken from the literature~\cite{azom}.

\begin{figure}[t]
\centering
\includegraphics[width=0.47\linewidth]{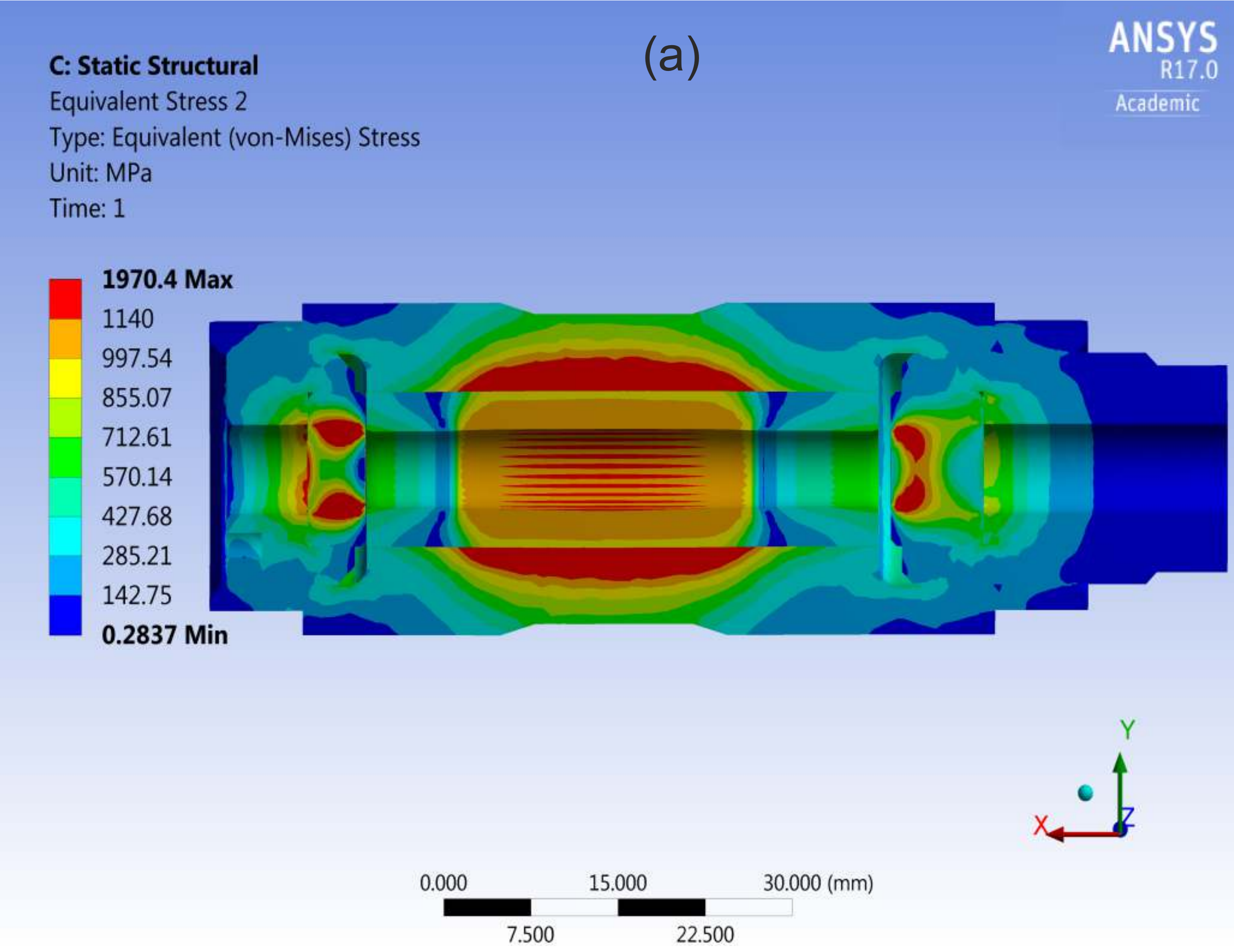}~~~~
\includegraphics[width=0.49\linewidth]{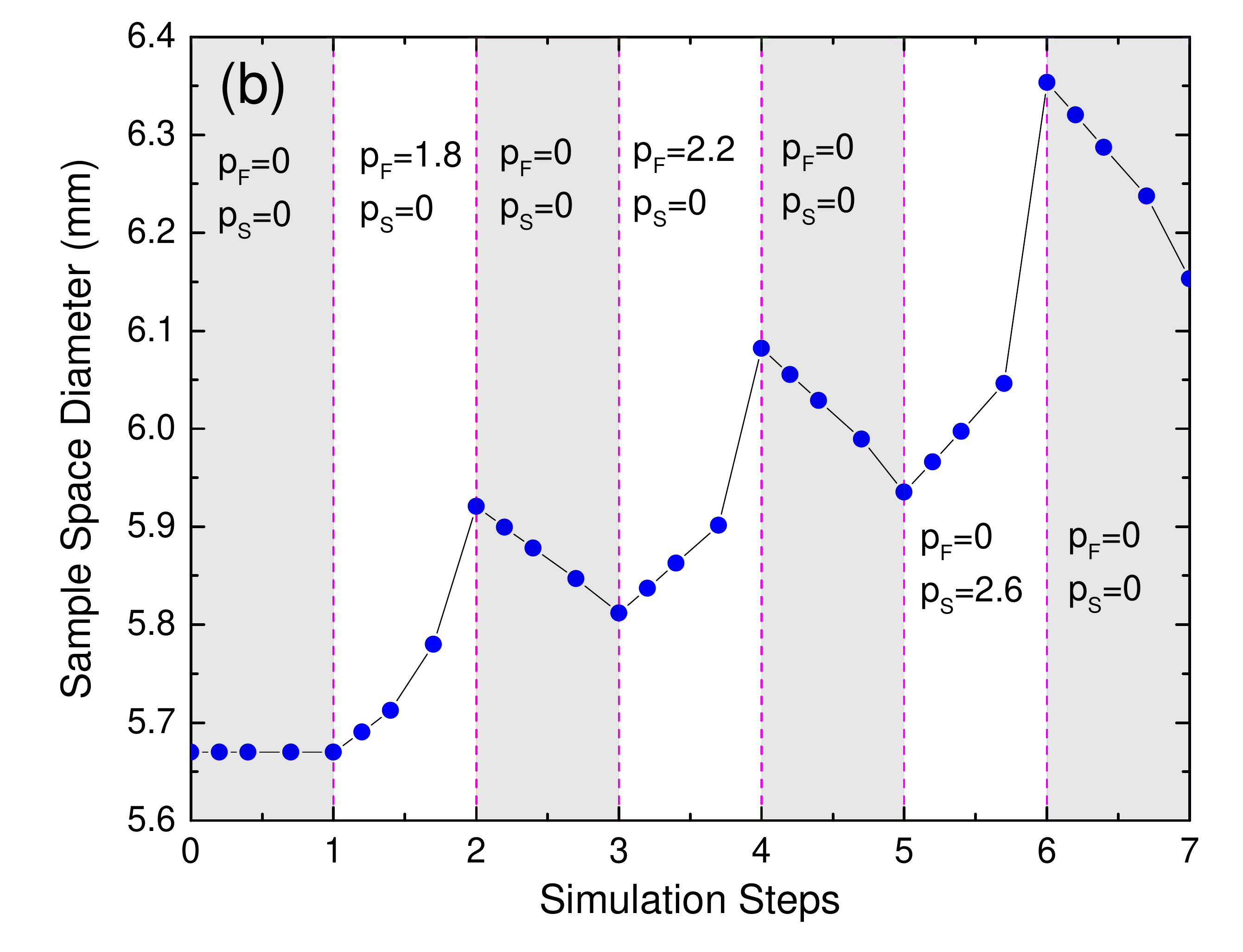}
%\vspace{-7.5cm}
\caption{\footnotesize (a) Equivalent (von-Mises) stress distribution throughout the cell. (b) Simulation of inner channel expansion. $p\rm_{F}$ and $p\rm_{S}$ denote the frettage and sample pressures, respectively.}
\label{Fig:VonMises}
\end{figure}

The optimization procedure consisted of three steps. As a first step, an analysis was performed in order to find the optimal mesh size and to choose the correct boundary conditions, including the contact behaviors. Once the preliminary analysis was satisfied (see Figure~\ref{Fig:VonMises}~(a)), the optimization using the Response Surface Optimization method was performed. This procedure allowed to estimate the basic range of parameters and to search for their possible correlations.  As a last step, the Goal Driven Optimization (GDO) method was applied. For better flexibility, 3D CAD software Autodesk Inventor~\cite{Inventor} was used as a designer software and for optimization. The communication between Inventor and ANSYS allowed performing the optimization procedure in fully automatic mode. Note that, GDO is an iterative multi-objective genetic algorithm calculating global maxima/minima of input/output parameters. It generates set of samples (typically 100), by further simulating them in the Static Structural block. As a result, input and output parameters were stored for further analysis, and the most promising candidates of input parameters were selected.

For the double-wall pressure cell analysis, the dimensions such as the height of the cell ($\simeq 75$~mm), the external and internal diameters ($\varnothing_{\rm ext} =24$~mm, $\varnothing_{\rm int} =6$~mm), the inner channel height (44~mm) and the sample space (24~mm) were kept constant considering the size limitations of $\mu$SR cryostats, available muon energies at $\mu$E1 beam line at PSI and the possibility of the muon-beam collimation. The external diameter of CuBe cylinder and the diameter difference between the inner (CuBe) and outer (MP35N) cylinders were optimized in order to maximize the pressure inside the cell by keeping the strains
developed in the materials within the limit of $\varepsilon_{\rm CuBe} \leq $ 0.07 and $\varepsilon_{\rm MP35N} \leq $ 0.04 mm/mm, respectively (red lines at Figure~\ref{Fig:StressStrain}).

 Results of the three step optimization of the double-wall pressure cell are presented in Tab.~\ref{Fig:Candidates}.  During the optimization, the applied pressure was kept within 1.5-2.0 GPa range in order to avoid appearance of overstrained areas caused by the change of cell dimensions, which might be the source of convergence problems. The optimization process resulted in three 'Candidate Points' with the corresponding radius of CuBe cylinder 5.7673, 5.1445, and 5.0156~mm, respectively (see Tab.~\ref{Fig:Candidates}).  To follow the low-background pressure cell concept described in the introductory section (see also Figure~\ref{Fig:SRIM}) the outer diameter of the inner CuBe cylinder had to be as big as possible.  For this reason the 'Candidate Point 1' with the outer radius of 5.75~mm and the overlapping difference of 70~$\mu$m was selected (Tab.~\ref{Fig:Candidates}).

For frettage simulation, the permanent deformation model, including the kinematic hardening model, was used. The initial diameter of the inner channel of CuBe cylinder was set to 5.65 mm. The full process was divided into seven steps as presented in Figure~\ref{Fig:VonMises}~(b). At the first step, there was no applied pressure. This allowed to capture interference contacts between cylinders and ANSYS interpreted it as a contact pressure. Second step corresponded to application of the 'frettage` pressure ($p_{\rm F}=1.8$~GPa) throughout the full pressure channel. At the third step the cell relaxed ($p_{\rm F}=0$) by introducing permanent deformation and remaining strains in the model. Note that in the fully relaxed state the inner diameter increased up to $\simeq 5.83$~mm. 2.2~GPa frettage pressure was applied for the step four and it was further removed at step five in order to relax the cell and to complete the imitation of frettage process. After step five the inner diameter of the cell reached $\simeq 5.95$~mm. Steps six and seven correspond to the application of 2.6~GPa pressure to the 'sample` region with the subsequent decreasing it to zero. At the end the permanent deformation of inner channel reached 6.15~mm, which is very close
to that measured experimentally~(see Section.~\ref{sec:cell-test}).

To conclude, the simulations presented above show that the maximum working pressure for the double-wall pressure cell with the parameters corresponding to the 'Candidate point 1` (see Tab.~\ref{Fig:Candidates}) is $\simeq 2.6$~GPa.
\begin{table}[t]
\begin{tabular}{l}
\includegraphics[width=1.0\linewidth]{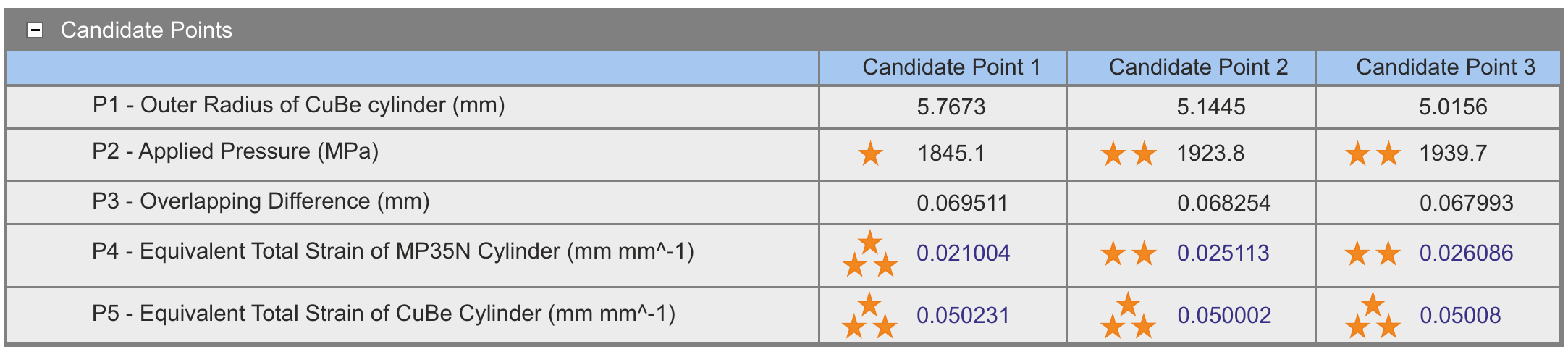}
\end{tabular}
\caption{\footnotesize Three candidate points obtained after FEA optimization. The parameters are: P1 - the outer radius of CuBe cylinder in mm; P2 -  the maximum holded pressure in MPa; P3 - the difference between the outer diameter of the inner cylinder and the inner diameter of the outer cylinder in mm; P4 - equivalent total strain of MP35N sleeve; P5 - equivalent total strain of inner CuBe cylinder; P4 and P5 are equivalent total strains of the outer (MP35N) and the inner (CuBe) cylinder, respectively. The last two parameters can not exceed the maximum strain limit set for CuBe and MP35N alloys (see Figure~\ref{Fig:StressStrain}) and they were minimized during the optimization procedure.  A three-star rating is provided by ANSYS and represents the goodness of simulations of the GDO for each candidate points.}
\label{Fig:Candidates}
\end{table}

\newpage

\section{Pressure cell design and assembly details} \label{sec:design_and_assembly}
%Cell design\\

\begin{figure}[t]
\centering
\includegraphics[width=0.32\linewidth]{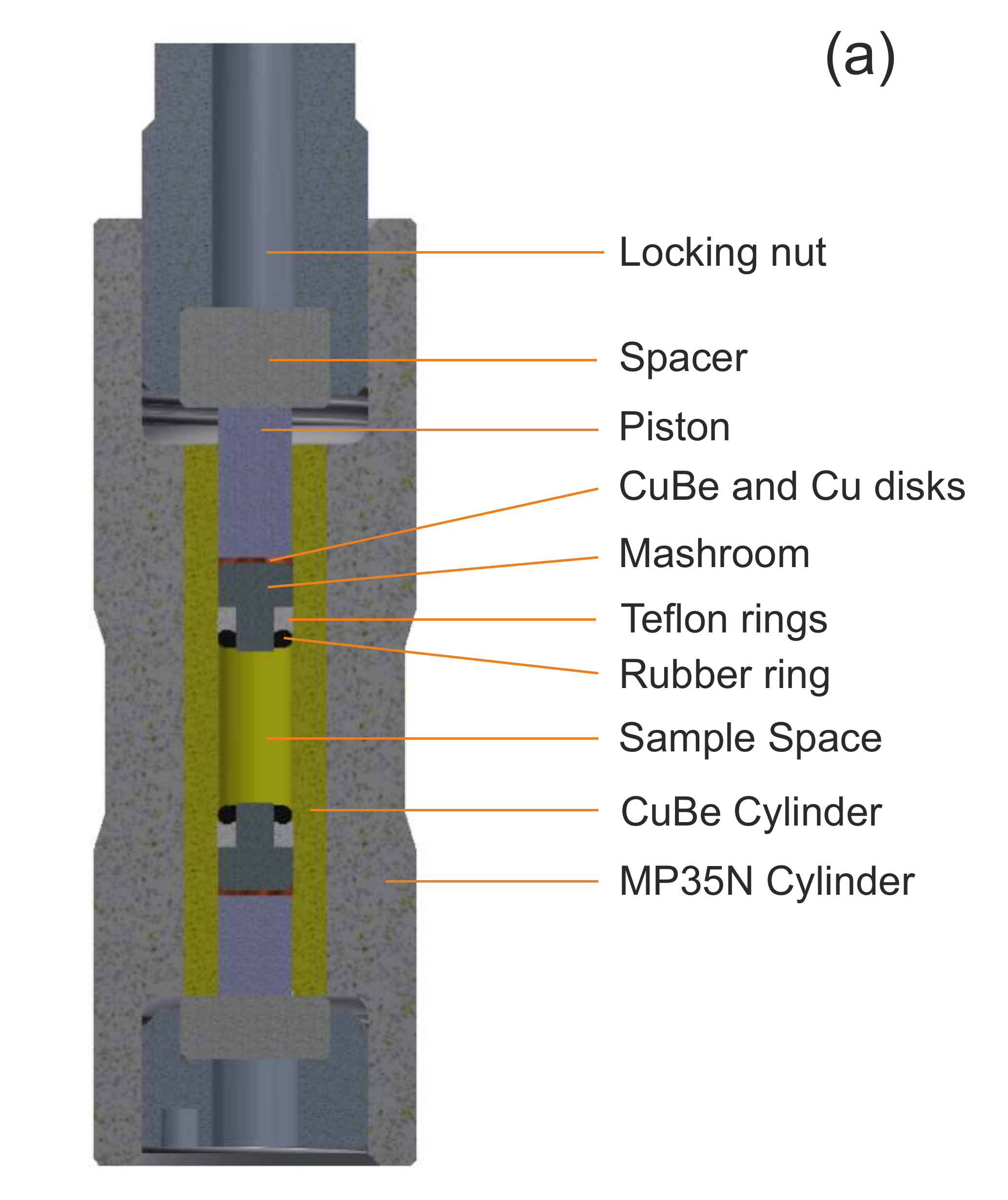}
\includegraphics[width=0.40\linewidth]{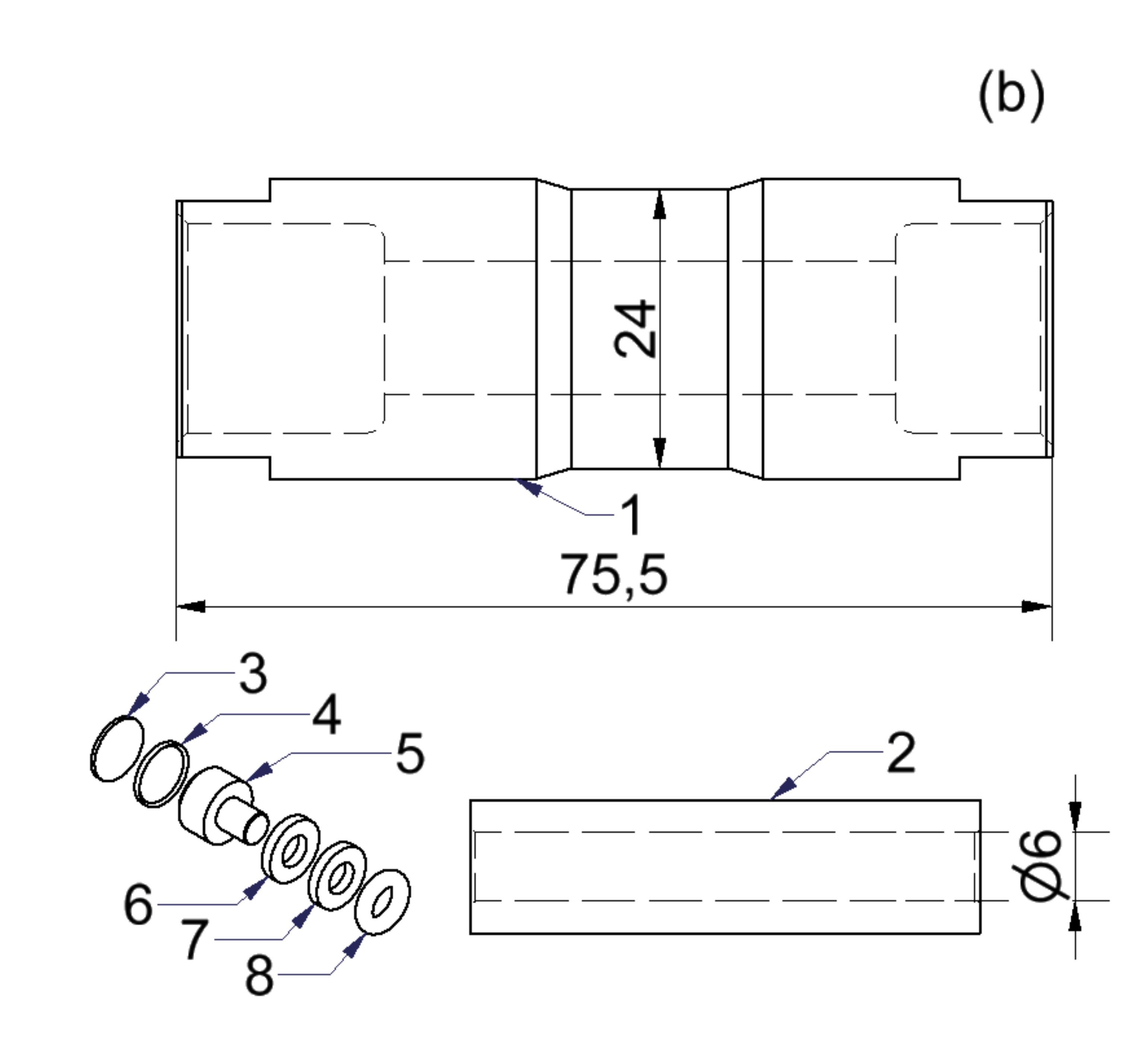}~~~~~
\includegraphics[width=0.22\linewidth]{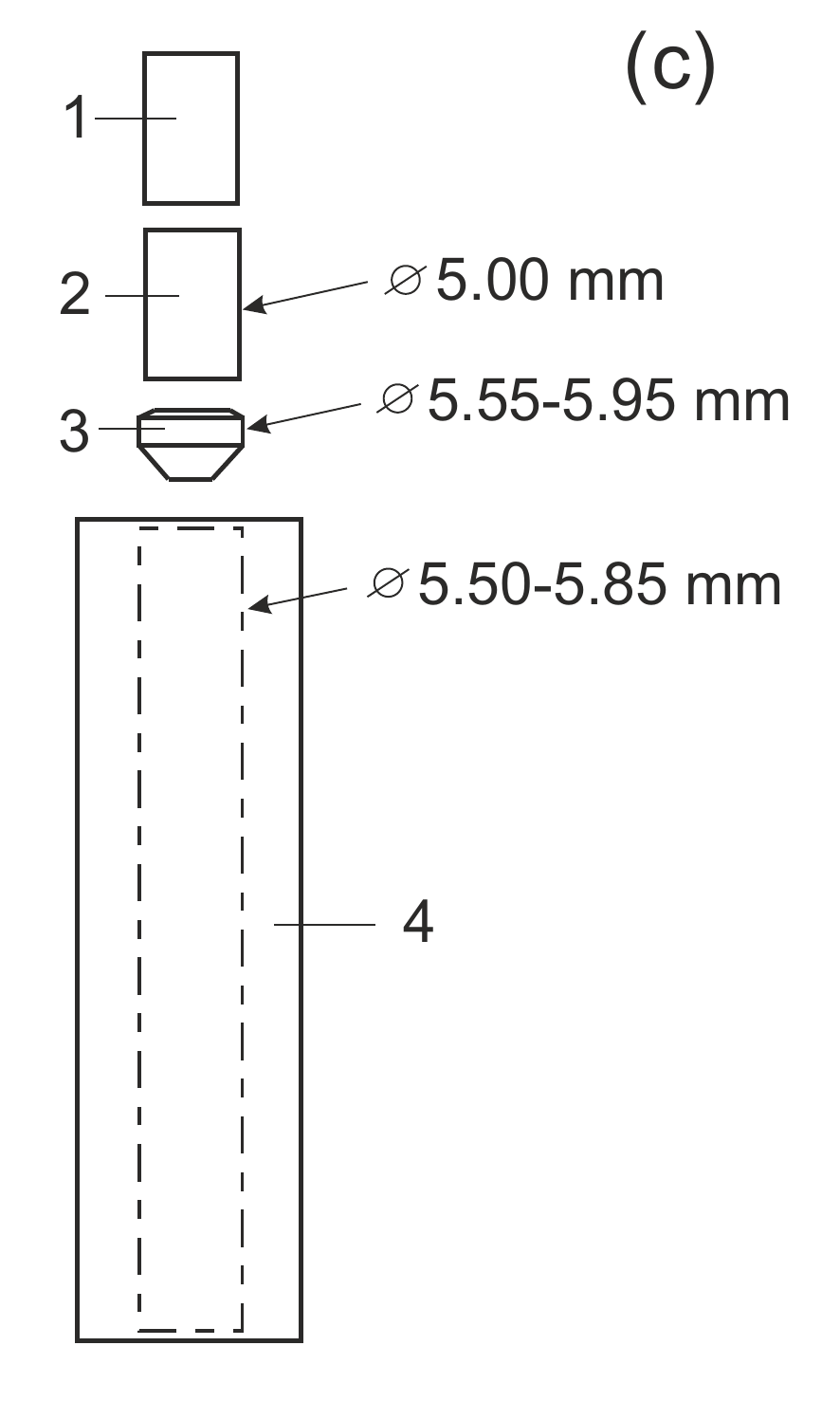}
%\vspace{-7.5cm}
\caption{\footnotesize  (a) Cross-sectional view of the double-wall hybrid pressure cell. (b) Pressure cell parts: 1 - outer MP35N sleeve; 2 - inner CuBe cylinder; 3 - CuBe disk; 4 - Copper disk; 5 - CuBe mushroom; 6 and 7 - teflon rings; 8 - rubber o-ring. (c) Schematic view of the frettage procedure. Only the inner CuBe cylinder is shown. 1 and 2 - pushing pistons; 3 - plunger; 4 - CuBe cylinder. The starting inner diameter of CuBe cylinder is 5.5~mm. By pushing the series of plungers with increased diameter (from 5.55 to 5.95~mm with 0.05 mm step) the diameter of CuBe cylinder expands up to 5.85~mm, and finally by rimming procedure reaches 6-6.05~mm.}
\label{Fig:PcellFull}
\end{figure}

The double-wall hybrid pressure cell constructed in accordance with the results of FEA simulations is depicted in Figure~\ref{Fig:PcellFull}. The outer body of the cell has a cylindrical shape and is made of MP35N alloy (Ni-Co-Cr-Mo-C)~\cite{cartech}. To strengthen parts of the cell staying outside of the muon beam, the top and the bottom diameters of the outer MP35N cylinder were increased up to 26~mm (Figure~\ref{Fig:PcellFull}~(b)). The inner cylinder represents CuBe alloy (Berylco-25)~\cite{Alloy25} fitted inside the outer sleeve, and its outer diameter is obtained from optimization method (see Section.~\ref{sec:FEA}).
The machined cylinders were hardened using temperature annealing procedure (CuBe -- 3~h at 315$^{\rm o}$C air cooled and MP35N -- 4 hours at 590$^{\rm o}$C air cooled, \cite{Annealing}). Before assembly the inner diameter of the outer sleeve and the outer diameter of the inner cylinder were equal, thus giving a possibility to assemble them with a small force. After fitting cylinders into each other the frettage of the cell was performed. The series of plungers with continuously increased diameter (0.05~mm in step) were pushed one after another through the inner channel of the cell. This expanded the inner and the outer diameter of the CuBe cylinder and, as a result, generated an interface pressure between  the sleeve and the inner body of the cell. To achieve a necessary interface pressure between the cylinders, first the expansion of the outer cylinder as a function of the interface pressure was simulated by FEA, and then by monitoring the expansion of the outer diameter of the MP35N cylinder during the frettage procedure, the desired interface pressure was achieved.

Considering that pistons stay relatively close to the sample space and, therefore, might be hit by the muon beam, a special search for the new low-background materials was performed. As possible candidates ZrO$_2$, Si$_3$N$_4$ and 'pure' WC (Kennametal, Ref.~\cite{kennametal}) materials were tested. All three candidates were checked at the muon beam and were found to show much weaker temperature and field responses as compared to 'nonmagnetic` WC (Ceratizit-CTF21R, Ref.~\cite{ceratizit}) used previously. The check of corresponding pistons inside the test pressure cell revealed that Si$_3$N$_4$ survives up to pressures of 3.0-3.2~GPa, while both ZrO$_2$ and WC (Kennametal) break much earlier.

The pressure cell seal system is of the similar type as described previously \cite{Khasanov1}. It consists of a 'mushroom', teflon rings and rubber o-ring. The rubber o-ring seals the system at low pressures. The teflon, which flows under the pressure by filling the space between the 'mushroom' and the pressure cell walls, works as a high-pressure seal.  In order to use Si$_3$N$_4$ piston with our 'mushroom' type sealing system the conical CuBe ring used previously \cite{Khasanov1} was replaced with the double-disk system consisting of 0.4~mm thick flat CuBe and conically shaped Cu disks (pieces 3 and 4 in Figure~\ref{Fig:PcellFull}~(b)). The use of disks compared to the conical ring reduces high stresses on ceramic pistons by evenly distributing the forces. Note that the main purpose of this part of the seal is to prevent the teflon to flow between the piston and the pressure cell walls.
The rest, spacers, top and bottom locking nuts, are made out of MP35N material, and does not participate to the $\mu$SR background signal as they located far away from the sample.  The drawing with corresponding dimensions of outer and inner cylinders in millimeters is depicted in Figure~\ref{Fig:PcellFull}~(b).

\section{Tests of the double-wall CuBe/MP35N cell} \label{sec:cell-test}

\begin{figure}[b]
	\centering
	\includegraphics[width=0.47\linewidth]{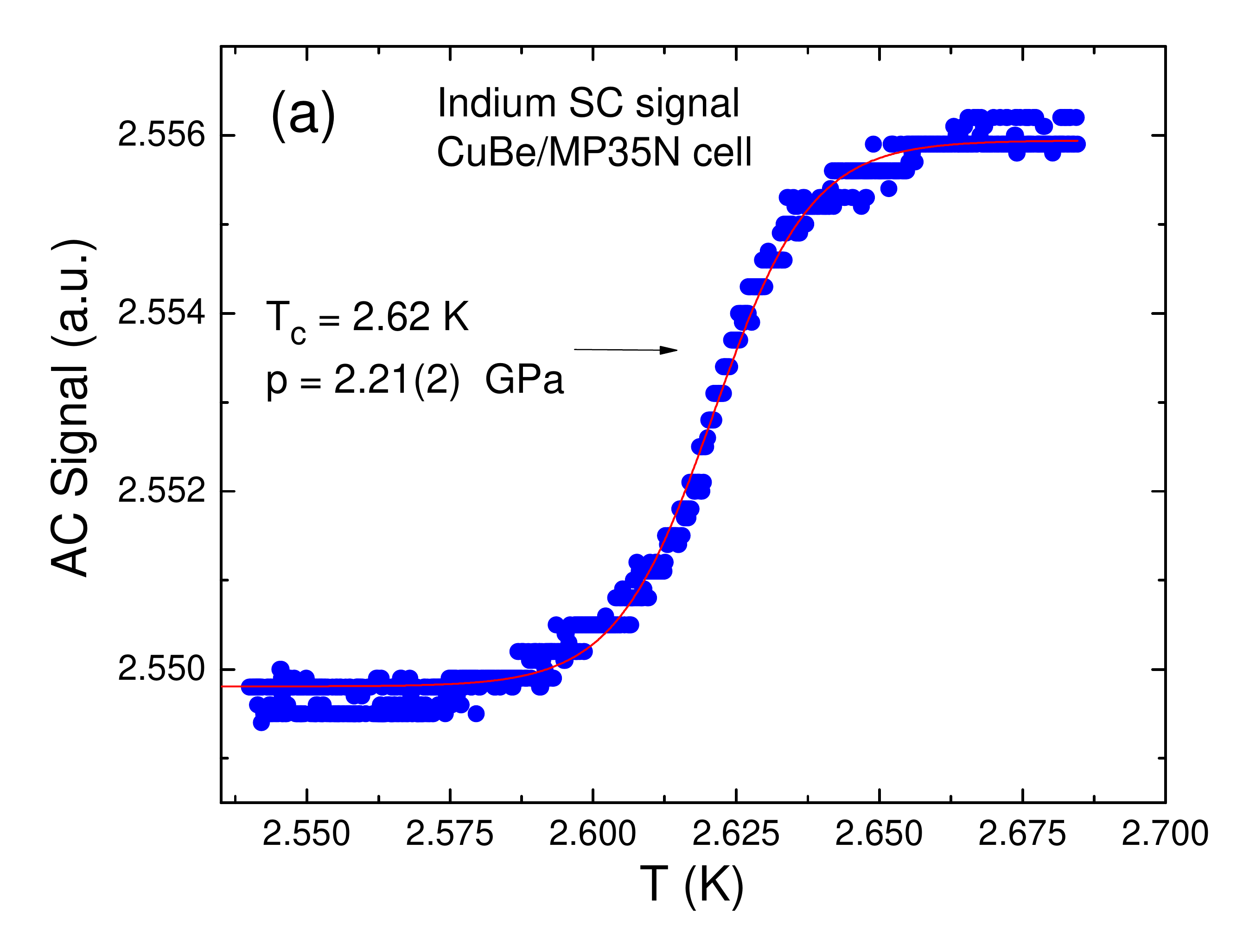}~~~~
	\includegraphics[width=0.45\linewidth]{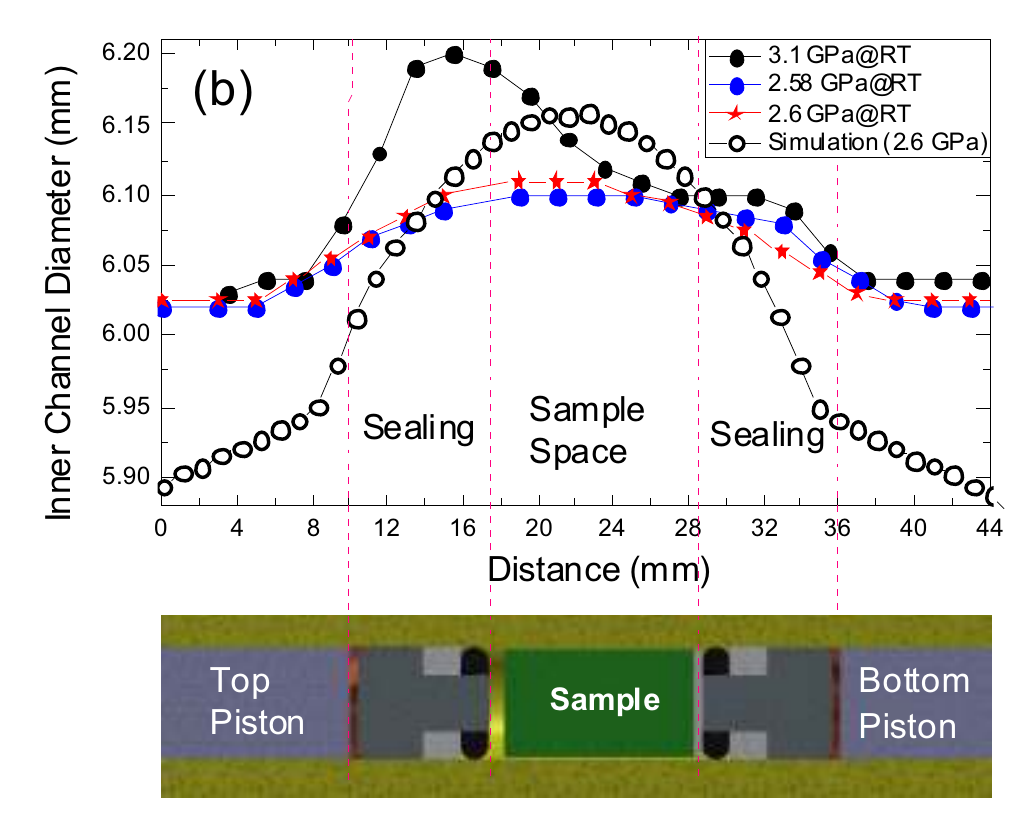}
	%\vspace{-7.5cm}
	\caption{\footnotesize (a) The superconducting transition of the piece of Indium (pressure indicator) measured by means of AC susceptibility. 2.6~GPa pressure is applied at room temperature. 2.21(2)~GPa is determined from the pressure induced shift of $T_{\rm c}$ of Indium indicator. (b) The expansion of the inner channel after pressure application.  }
	\label{Fig:TcandDexp}
\end{figure}

\begin{figure}[t]
	\centering
	\includegraphics[width=0.326\linewidth]{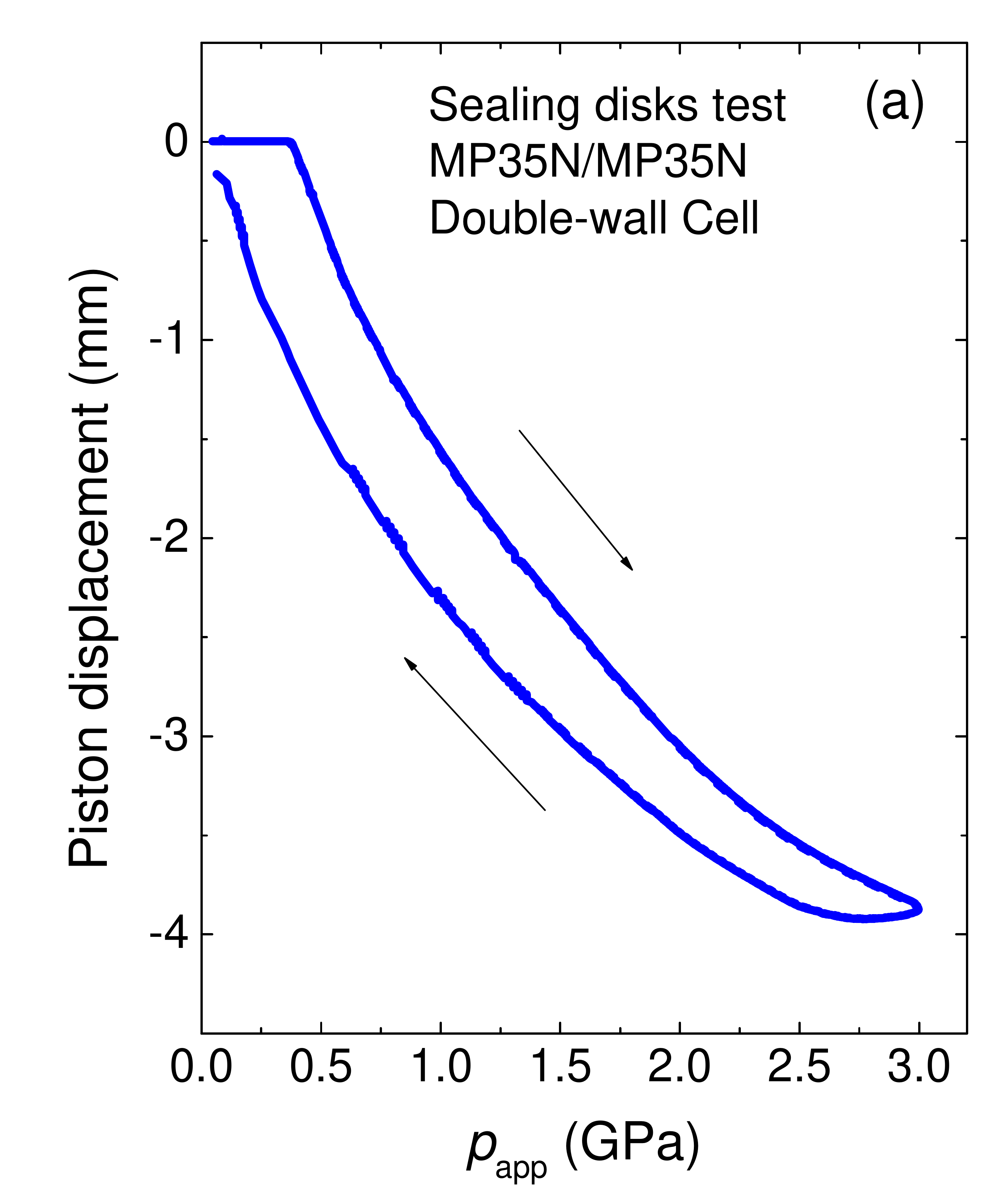}
	\includegraphics[width=0.326\linewidth]{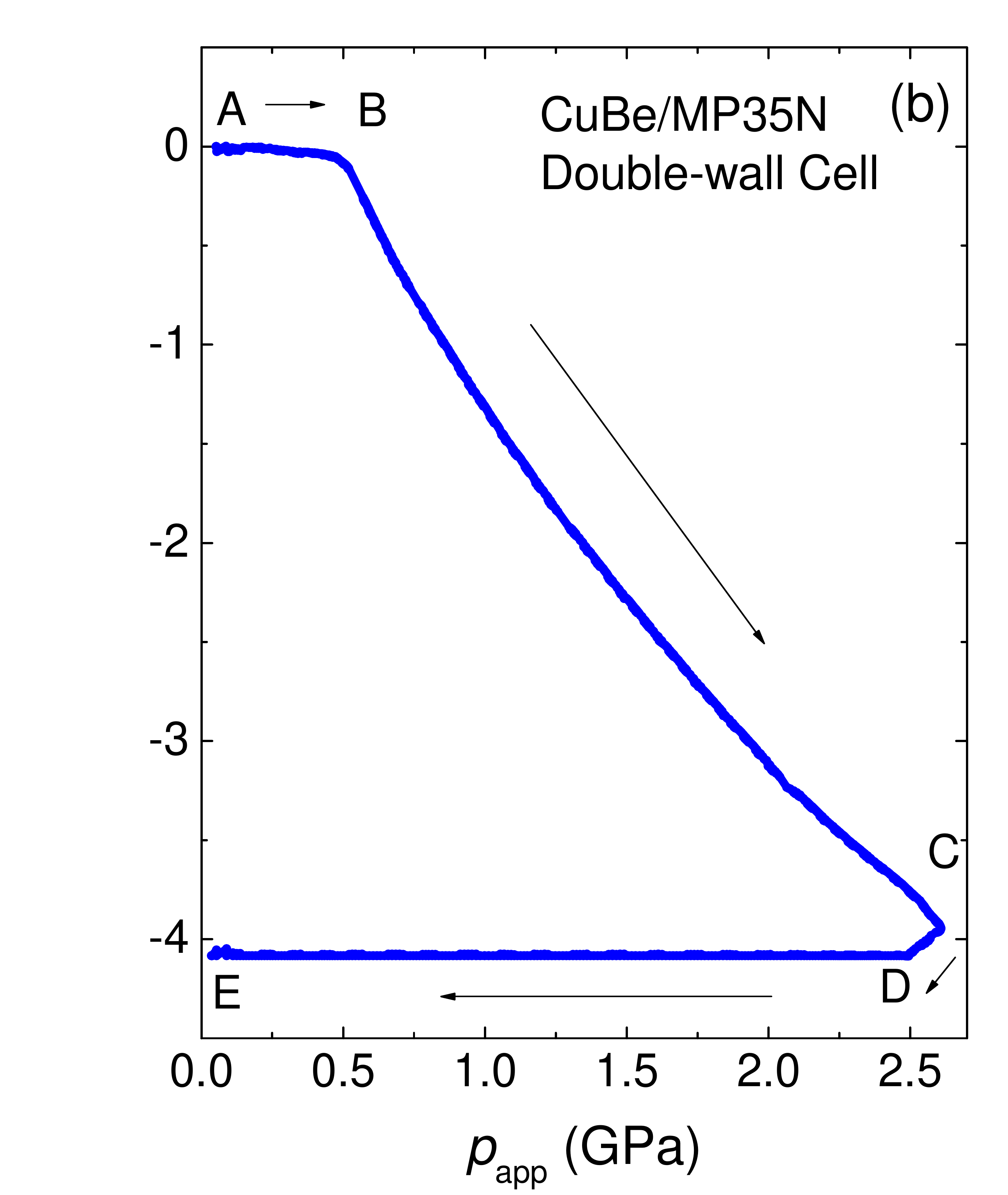}
	\includegraphics[width=0.326\linewidth]{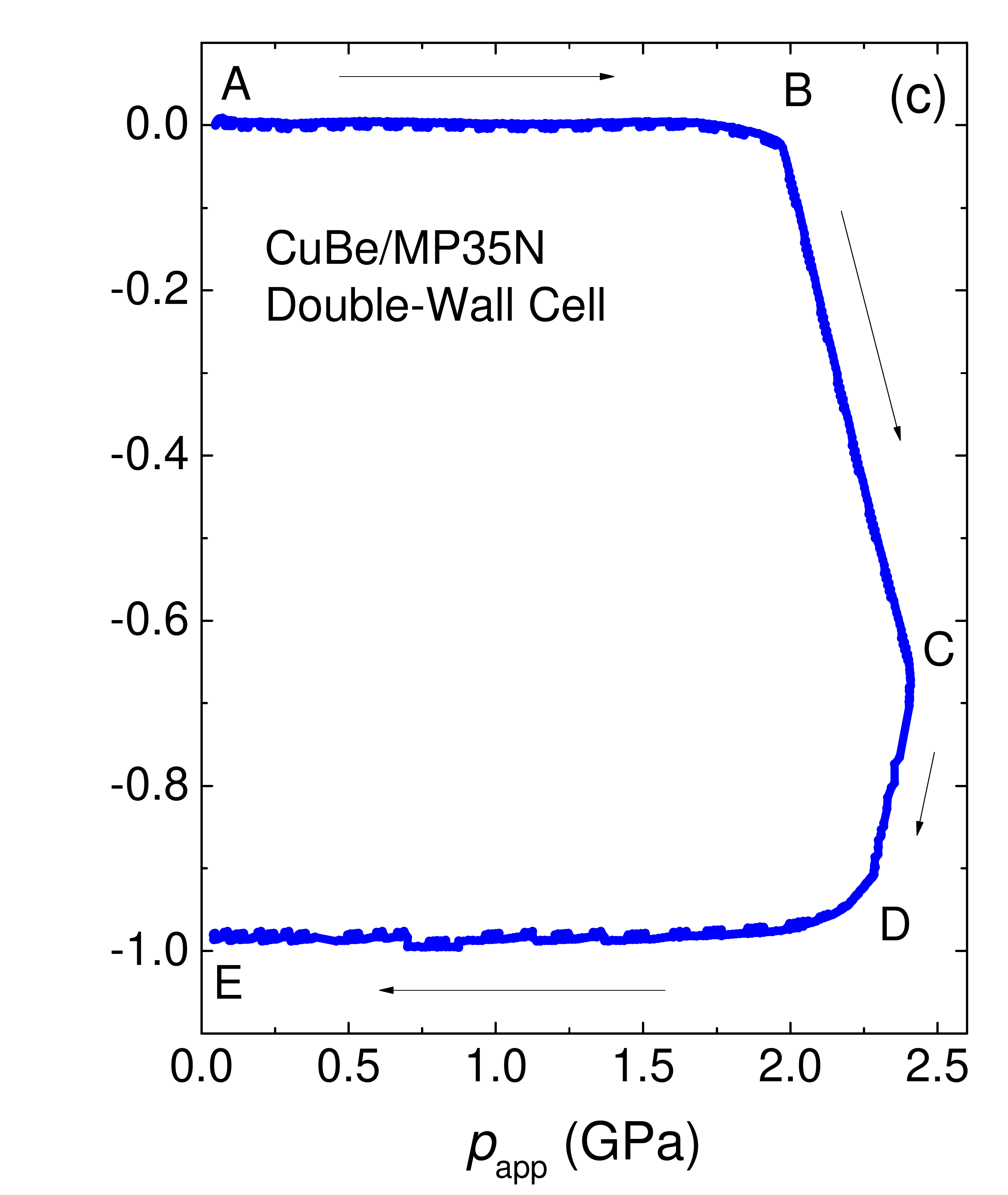}
	%\vspace{-7.5cm}
	\caption{\footnotesize Piston displacement as a function of applied pressure at room temperatures. (a) The test of 'double-disk` sealing system. The difference between the up and down curve corresponds to twice the friction within the sealing system (see text for details). (b) The process of closing the double-wall CuBe/MP35N pressure cell. The baranches correspond to: A$\rightarrow$B -- the piston does not move (the pressure inside the cell ($\simeq 0.5$~GPa) is caused by the initial preseal); B$\rightarrow$C -- the piston moves inside the cell by compressing the pressure transmitting medium; C$\rightarrow$D -- closing the cell by tightening the top locking nut; D$\rightarrow$E -- releasing the applied pressure to zero (the piston remains supported by the spacer disk and the pressure cell becomes completely sealed). Consequently, the decrease of $p_{\rm app}$ from point D to E do not leave to any piston movement. (c) The process of pressure change from $\simeq 2.0$ to $\simeq 2.4$~GPa. (A$\rightarrow$B; B$\rightarrow$C; C$\rightarrow$D and D$\rightarrow$E paths have similar meaning as in panel b).}
	\label{Fig:PistDispl}
\end{figure}

\begin{figure}[b]
\centering
\includegraphics[width=0.49\linewidth]{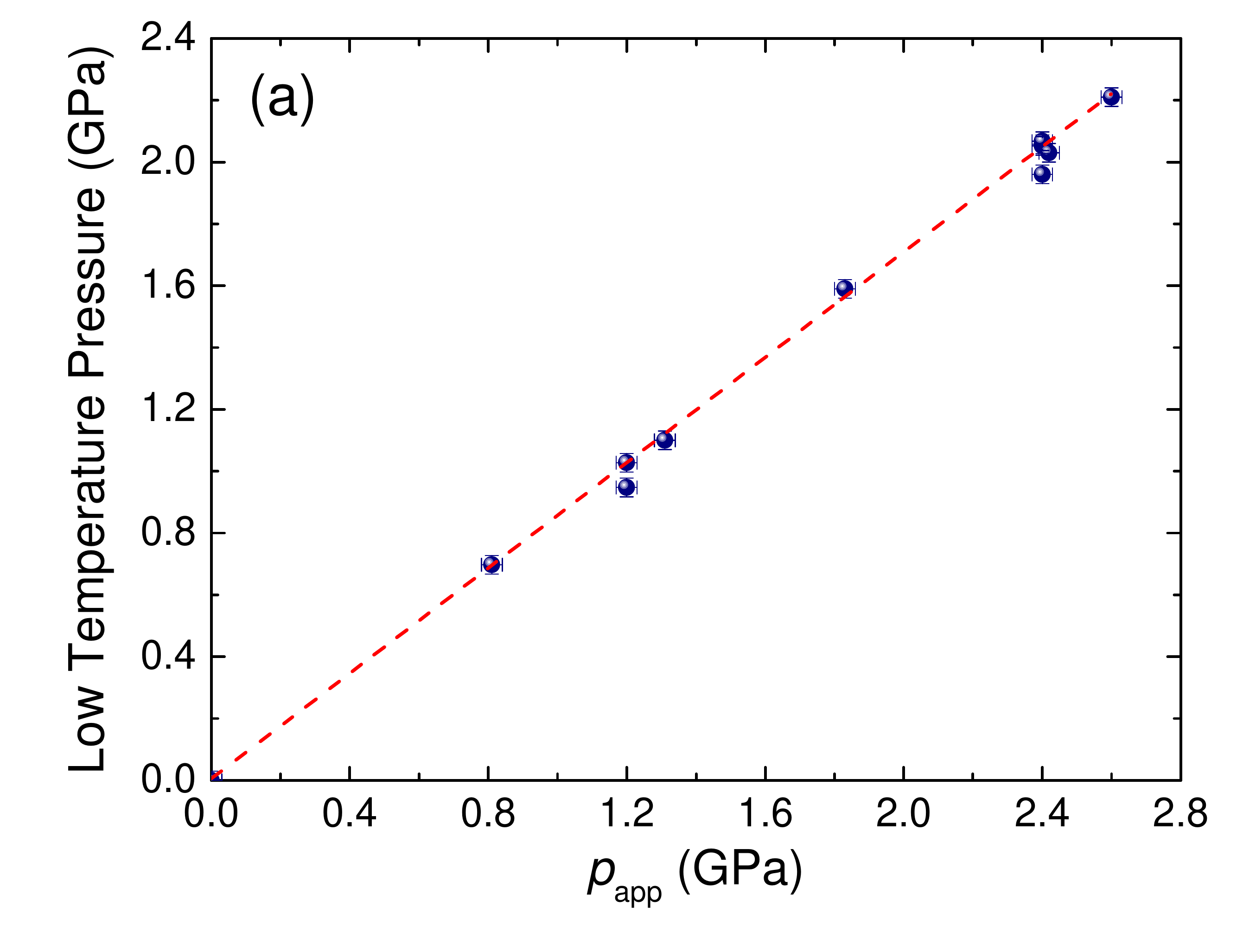}
\includegraphics[width=0.49\linewidth]{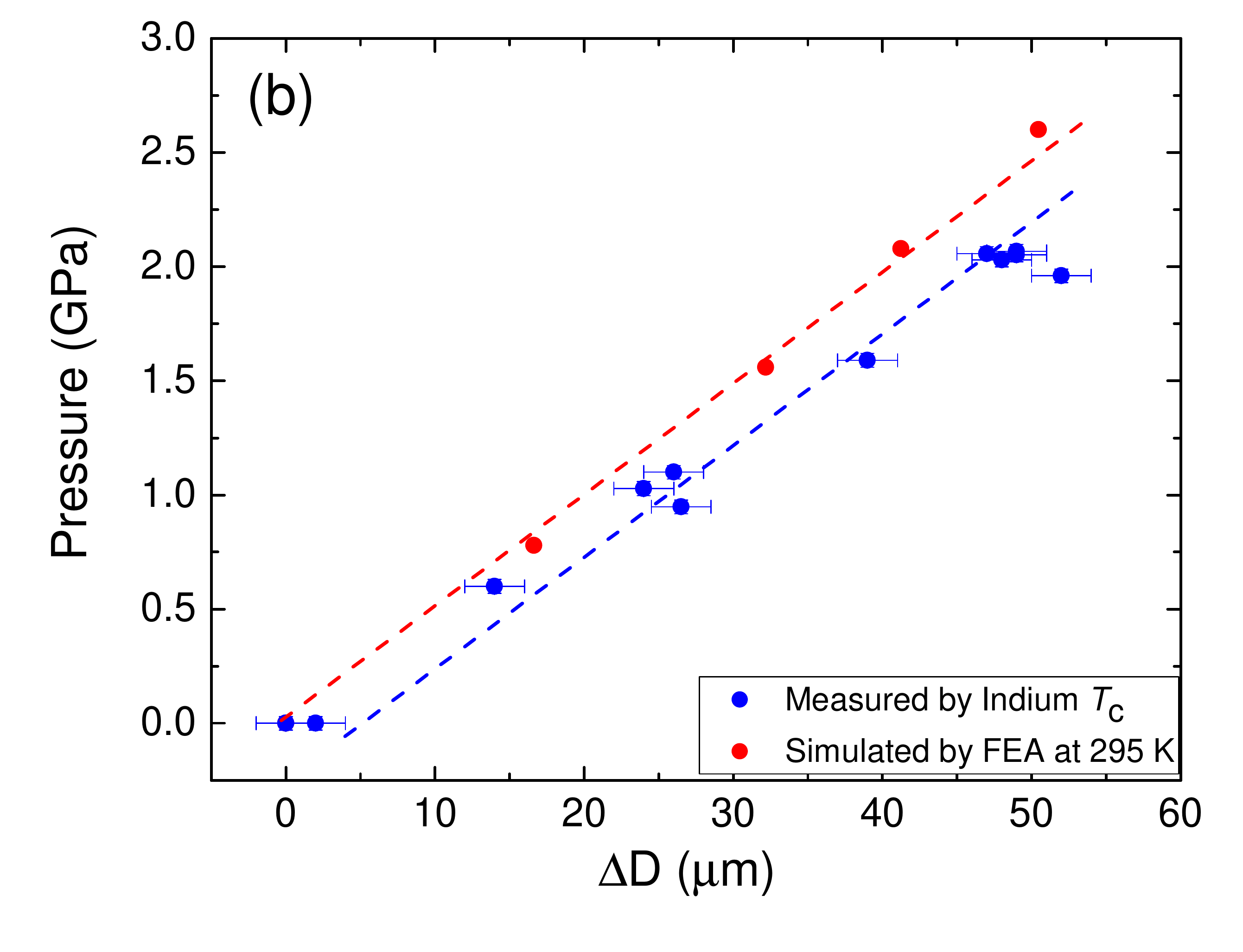}
%\vspace{-7.5cm}
\caption{\footnotesize (a) Measured low temperatures pressure as a function of applied pressure at room temperature; (b) The simulated (red dots) and the measured (blue squares) pressure inside the cell as a function of the change of diameter of the pressure cell $\Delta {\rm D}$. The lines are guides to the eye.}
\label{Fig:OuterDexp}
\end{figure}

The pressure cell was tested by pressurizing it several times. Pressure inside the cell was measured by monitoring the superconducting transition temperature ($T_{\rm c}$) of metallic Indium (pressure indicator) placed inside the cell (see Figure~\ref{Fig:TcandDexp}~a and Ref.~\cite{Khasanov1} for details). The low-temperature pressure value and the expansion of the outer diameter of the cell was measured after each pressure application. The expansion of the inner channel was studied at the end of every pressure application cycle with the highest pressure reached 3.1/2.6/2.58~GPa in cycle 1/2/3, respectively.
 The corresponding dependence of the inner diameter of the inner bore as a function of distance from the top of the inner cylinder is depicted in Figure~\ref{Fig:TcandDexp}~(b). Two important points need to be mentioned. (i) The 'room temperature' (RT) pressure value (applied pressure value $p_{\rm app}$) shown in Figs.~\ref{Fig:TcandDexp}~(b), \ref{Fig:PistDispl}, and \ref{Fig:OuterDexp}) is obtained as $p_{\rm app}=F/S$ ($F$ is the force applied by the press and $S$ is the area of the sample channel) and does not take into account the effects of friction within the sealing system \cite{Walker_99}. (ii) After applying the load which would have theoretically generated the pressure of 3.1~GPa the inner channel near the sample space was barreled so strongly (black solid symbols at Figure~\ref{Fig:TcandDexp}~(b)) that the sealing system was no longer capable of holding the pressure. The inner CuBe cylinder was drilled out and replaced with a new one. The experiments presented below were performed by using the second CuBe cylinder. Our experience showed that the maximum expansion of the inner channel after which the cell cannot be closed anymore is about 0.2~mm. Figure~\ref{Fig:TcandDexp}~(b) implies that the application of 2.6~GPa leads to the maximum expansion of $\simeq 0.1$~mm which is under the limit. Remarkably, the second cycle, with the maximum applied pressure 2.58~GPa, did not lead to any further expansion of the pressure cell channel. This means that pressures up to 2.6~GPa could be repeatedly and safely applied to the present cell. The low-temperature pressure value obtained from the measured indium superconducting transition is found to be $\simeq 2.2$~GPa (see Figure~\ref{Fig:TcandDexp}~(a)).


The difference between the high-temperature applied (2.6~GPa) and the low-temperature measured (2.2~GPa) pressure values originates from (i) a thermal shrinking of a pressure transmitter medium (Daphne 7373 oil in our case, Ref.~\cite{daphne}) and (ii) a friction within the pressure seal system. The first one arises from the difference in thermal expansion coefficients of the pressure cell body (including pistons and mushroom seals) and the pressure transmitting medium. The typical pressure drop inside the cell caused by cooling it down from the room temperature to $T\simeq 5$K is $\simeq 0.25$~GPa for Daphne 7373 oil \cite{Kamarad_04, Kamenev_06}. The friction, which is the second source of the low-temperature pressure drop, can be estimated by monitoring the cell load-unload cycle. Figure~\ref{Fig:PistDispl} shows a few examples of displacement of the top piston as a function of applied pressure. Figure~\ref{Fig:PistDispl}~(a) represents the case when the applied pressure was increased up to $\simeq 3.1$~GPa and further released to zero without closing the cell. This experiment was carried out in order to test the capability of the new sealing system (see Section.~\ref{sec:design_and_assembly} and Figure~\ref{Fig:PcellFull}~(b)) and was performed using our previously reported double-wall cell made completely out of MP35N alloy \cite{Khasanov1}. The hysteresis between the up and the down branches of the load-unload curve is caused by the friction within the sealing system. By increasing $p_{\rm app}$ the real pressure inside the cell needs to be lower than the applied one: $p_{\rm app}-F_{\rm fr}/S$ ($F_{\rm fr}$ is the friction force and $S$ is the area of the pressure channel), while by decreasing $p_{\rm app}$ the pressure inside the cell is higher than the applied one: $p_{\rm app}+F_{\rm fr}/S$. Consequently, the width of the hysteresis curve should correspond to twice the pressure loss within the sealing system. By following Figure~\ref{Fig:PistDispl}~(a) the difference between the loading and unloading branches ($\simeq0.35$~GPa) is independent on $p_{\rm app}$ and corresponds to the pressure loss within the sealing of $\simeq 0.18$~GPa. The process of closing and changing the pressure in CuBe/MP35N cell is presented in Figs.~\ref{Fig:PistDispl}~b and c. In panel b the cell, which is initially pre sealed by tightening the top locking nut (see Figure~\ref{Fig:PcellFull}~(a)), is closed at $p_{\rm app}\simeq 2.6$~GPa. The full process is described as follows. By increasing $p_{\rm app}$ from point A to B the piston does not move. The pressure inside the cell ($\simeq 0.5$~GPa) is caused by the initial pre seal. With the further $p_{\rm app}$ increase (region B$\rightarrow$C) the piston moves inside the cell by compressing the pressure transmitting medium. The region C$\rightarrow$D corresponds to the pressure cell closing.  $p_{\rm app}$ decreases and the piston goes further inside the cell due to tightening the locking nut (see Figure~\ref{Fig:PcellFull}~(a)). At the point D the piston is supported by the locking nut via the spacer disk and the pressure cell becomes completely sealed. Consequently, the decrease of $p_{\rm app}$ from the point D to E does not result in any additional piston movement. Figure~\ref{Fig:PistDispl}~(c) describes the process of the pressure increase from $\simeq 2.0$ to $\simeq2.4$~GPa. It should be noted here that the pressure cell load with the simultaneous measurement of the piston displacement curve allows us to keep the pressure cell under the full control. Any sudden change of the slope, the step like displacement of the piston or even slight bend of the loading curve towards lower pressure direction could correspond either to the leak in the seal system, braking the piston, or, which is even more dangerous, pointing to exceeding the elastic limit of the pressure cell material. In any of such cases one should immediately release the pressure, open the cell and try to solve the problem(s). In some cases the presence of small piece of Indium inside the cell becomes unpractical. The stray field form the superconducting In ($T_{\rm c}\simeq 3.40$~K at ambient pressure) may distort the $\mu$SR sample response. One needs therefore to search for another way(s) to determine the pressure inside the cell. The simplest way is to correlate the applied and the measured pressures. Figure~\ref{Fig:OuterDexp}~(a) shows that there is a good correspondence between them. Another way is to relate the pressure inside the cell to the expansion of the outer diameter (see e.g.~\cite{Diederichs1, Diederichs2, Kamenev_06}). They show that the expansion of the outer diameter of the pressure cell cylinder corresponds directly to the pressure inside the cell. In reference~\cite{Kamenev_06} it was shown that by increasing the pressure inside the CuBe cell (8.7~mm outer diameter and 3.0~mm inner diameter) from 0 to $\simeq 1$~GPa, the cell expands by $\simeq 16$~$\mu$m, which is easily measurable quantity. In order to apply this technique for CuBe/MP35N double-wall cell described here the expansion of the outer diameter as a function of applied pressure was first simulated by FEA technique and than compared with the experimentally measured values (see Figure~\ref{Fig:OuterDexp}~(b)). Red and blue dots correspond to results obtained by FEA (at $T=295$~K) and from the pressure induced $T_{\rm c}$ shift of In ($T_{\rm c} \lesssim 3.4$~K), respectively. The shift of $\simeq 0.25$~GPa between the measured and simulated points is determined by the temperature compression of the pressure transmitting medium, which is in agreement with the values reported in the literature \cite{Kamarad_04, Kamenev_06}.

\section{$\mu$SR studies of background contributions} \label{sec:background-cntributions}

\subsection{$\mu$SR background of the pressure cell}

 In order to analyse the $\mu$SR experimental data, it is important to know the $\mu$SR response of the  pressure cell. Therefore, separate zero-field (ZF) and transverse-field (TF) $\mu$SR measurements were dedicated to obtain background signal of the cell filled with CuBe cylindricaly shaped sample and Daphne 7373 oil. The experiments were preformed at the GPD instrument (PSI, Switzerland) in the temperature range form 0.25 to 20~K. In TF-$\mu$SR experiments the magnetic field (30~mT) was applied perpendicular to the initial muon-spin polarization.

The ZF-$\mu$SR data were found to be well described by the static Kubo-Toyabe function with an additional exponential relaxation term as:
  \begin{eqnarray}
P(t)_{\rm ZF}&=& \left[\frac{1}{3} + \frac{2}{3}(1-\sigma_{\rm GKT}^{2}t^{2})e^{-\sigma_{\rm GKT}^{2}t^{2}/2}  \right]e^{-\lambda_{\rm ZF}t}~.
\label{eq:PcellBGSZF}
\end{eqnarray}
Here $\sigma_{\rm GKT}$ and $\lambda_{\rm ZF}$ are Gaussian Kubo-Toyabe and exponential relaxations.

The TF-$\mu$SR data were analysed by combining Gaussian ($\sigma_{\rm TF}$) and simple exponential ($\lambda_{\rm TF}$) relaxation functions:
  \begin{eqnarray}
P(t)_{\rm TF}&=& e^{-\lambda_{\rm TF}t} e^{-\sigma_{\rm TF}^{2}t^{2}/2} \cos(\gamma_{\mu}B_{\rm ap}t+\varphi).
\label{eq:PcellBGSTF}
\end{eqnarray}
Here $\gamma_{\mu}=2\pi \; 135.5$~MHz/T is the muon gyromagnetic ratio, $B_{ap}$ is the applied field and $\varphi$ is the initial phase of the muon-spin ensemble.

\begin{figure}[h]
	\centering
	\includegraphics[width=0.49\linewidth]{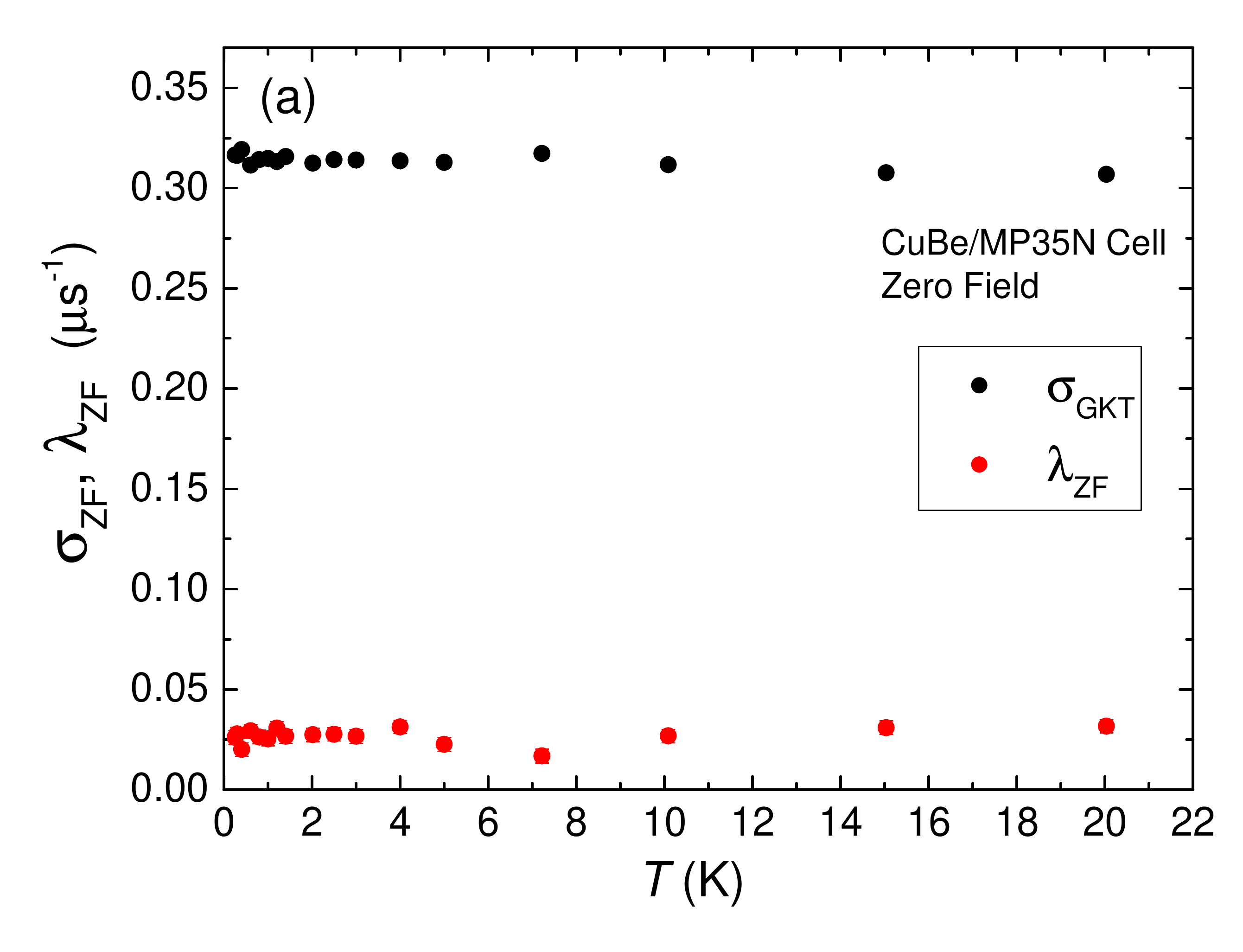}
	\includegraphics[width=0.49\linewidth]{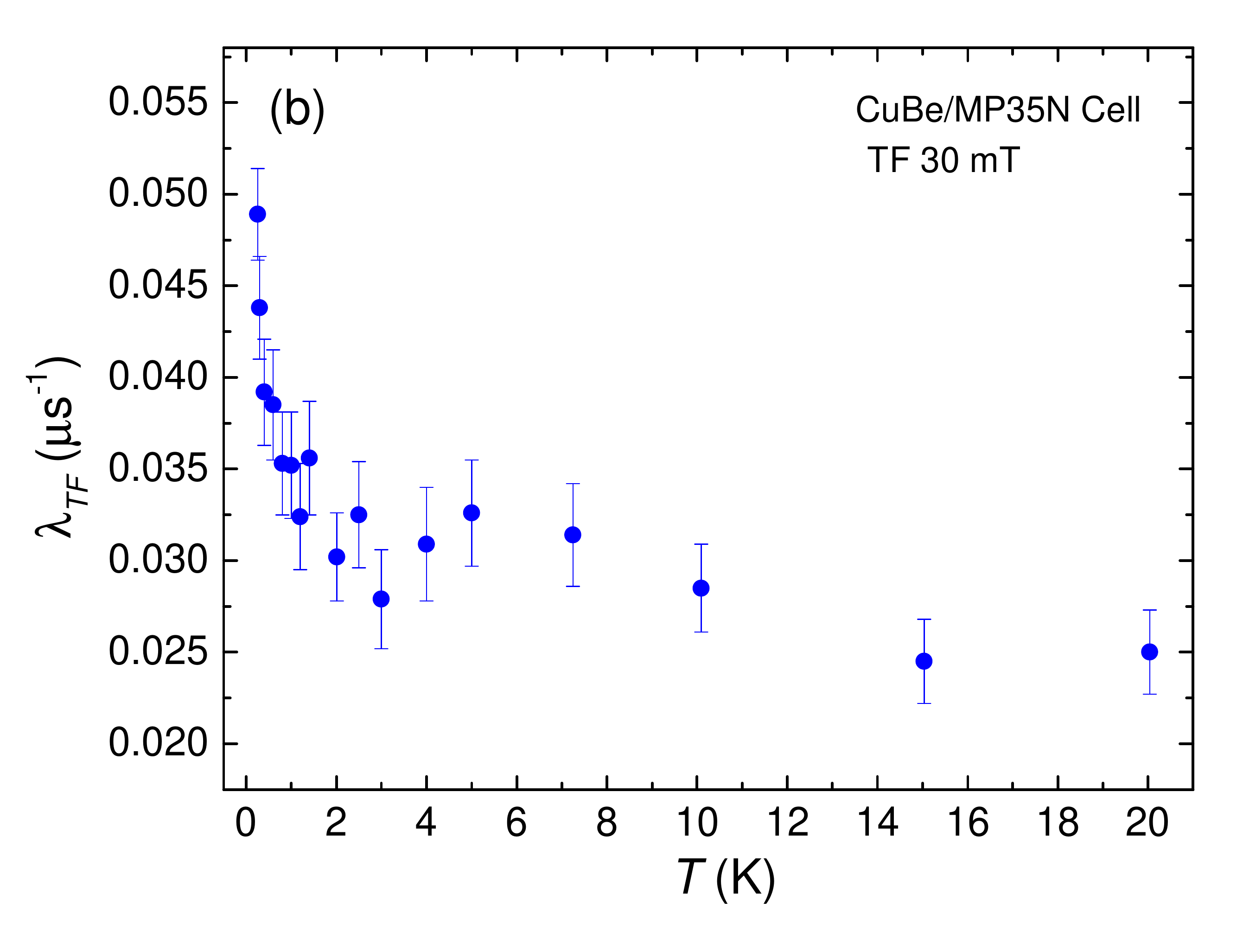}
	%\vspace{-7.5cm}
	\caption{\footnotesize Temperature dependencies of (a) ZF and (b) TF at 300~Gauss $\mu$SR relaxation rates in CuBe/MP35N double-wall cell.}
	\label{Fig:BGS}
\end{figure}

 Temperature dependencies of $\lambda_{\rm ZF}$, $\sigma_{\rm GKT}$ and $\lambda_{\rm TF}$ are presented in Figure~\ref{Fig:BGS}. Both ZF relaxation rates are temperature independent in 0.25-20~K range. Slightly increased  TF exponential rate was observed below 2~K, reaching highest 0.049 $\mu$s$^{-1}$ value at 250~mK. The gaussian component $\sigma_{\rm TF}$ of TF polarization function was found to the temperature independent in the $0.25-20$~K region and was 0.26 $\mu$s$^{-1}$.

  \subsection{$\mu$SR background of Si$_{3}$N$_{4}$ pistons}

In order to make true low-background pressure cell suitable for $\mu$SR studies, a special
attention needs to be paid to the material of compressing pistons. The reason is that,
by staying in close vicinity to the sample, pistons could be partially exposed to the
muon beam and thus, contribute to the background signal. By following their mechanical
properties and 'weak' temperature dependent $\mu$SR response the  Silicon Nitride was selected \cite{Si3N4}. It is a chemical compound consisting of silicon and nitrogen, with the formula Si$_3$N$_4$, and was originally developed for automotive engine wear parts.
Due to its advanced mechanical properties, Si$_{3}$N$_{4}$ is used in many severe mechanical, thermal and wear applications. High-quality parts developed for these demanding applications, with Young's Modulus close to 300~GPa and compressive strength up to 5500~MPa~\cite{Si3N4}, are available on the market today.

  \begin{figure}[b]
	\centering
	\includegraphics[width=0.49\linewidth]{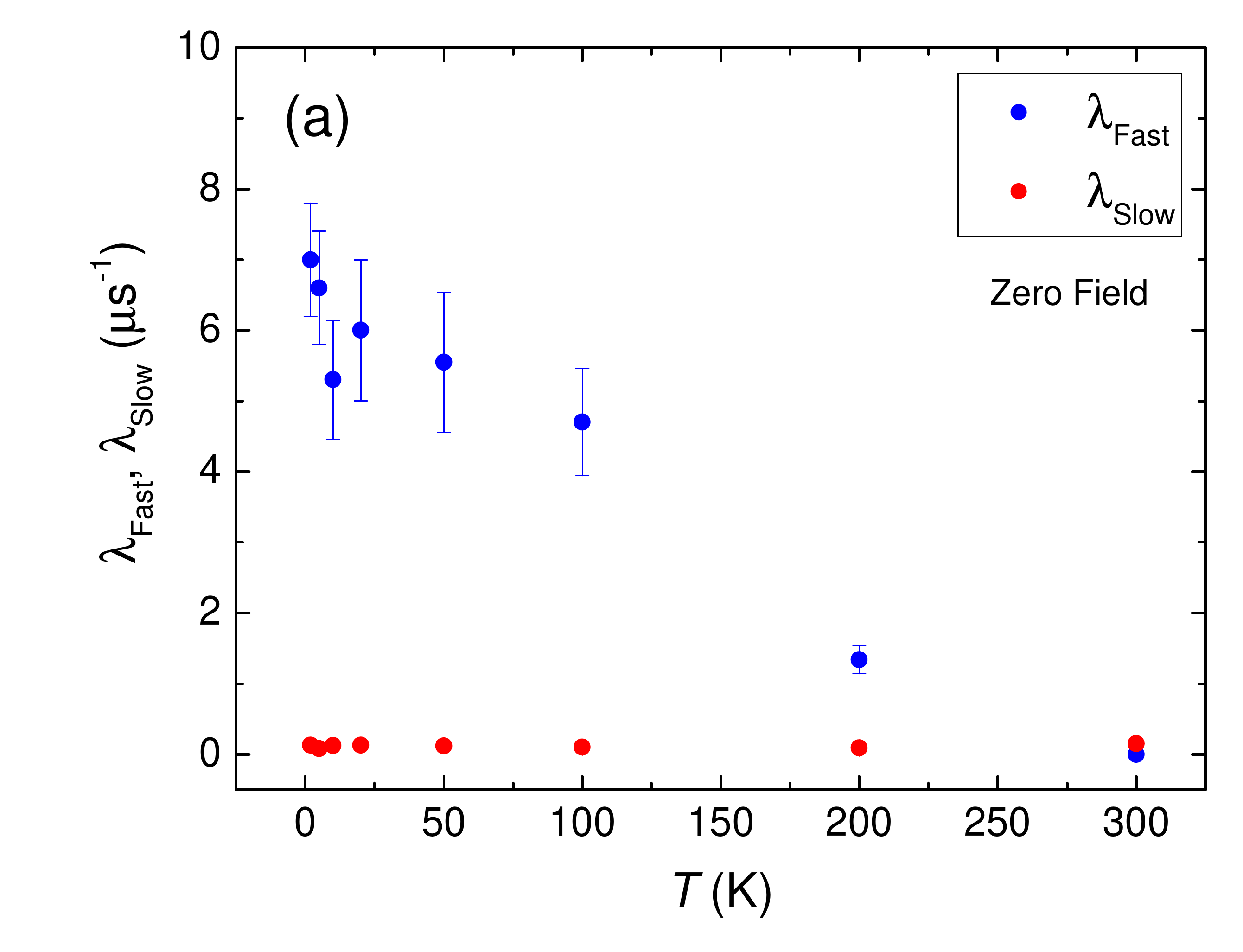}
	\includegraphics[width=0.49\linewidth]{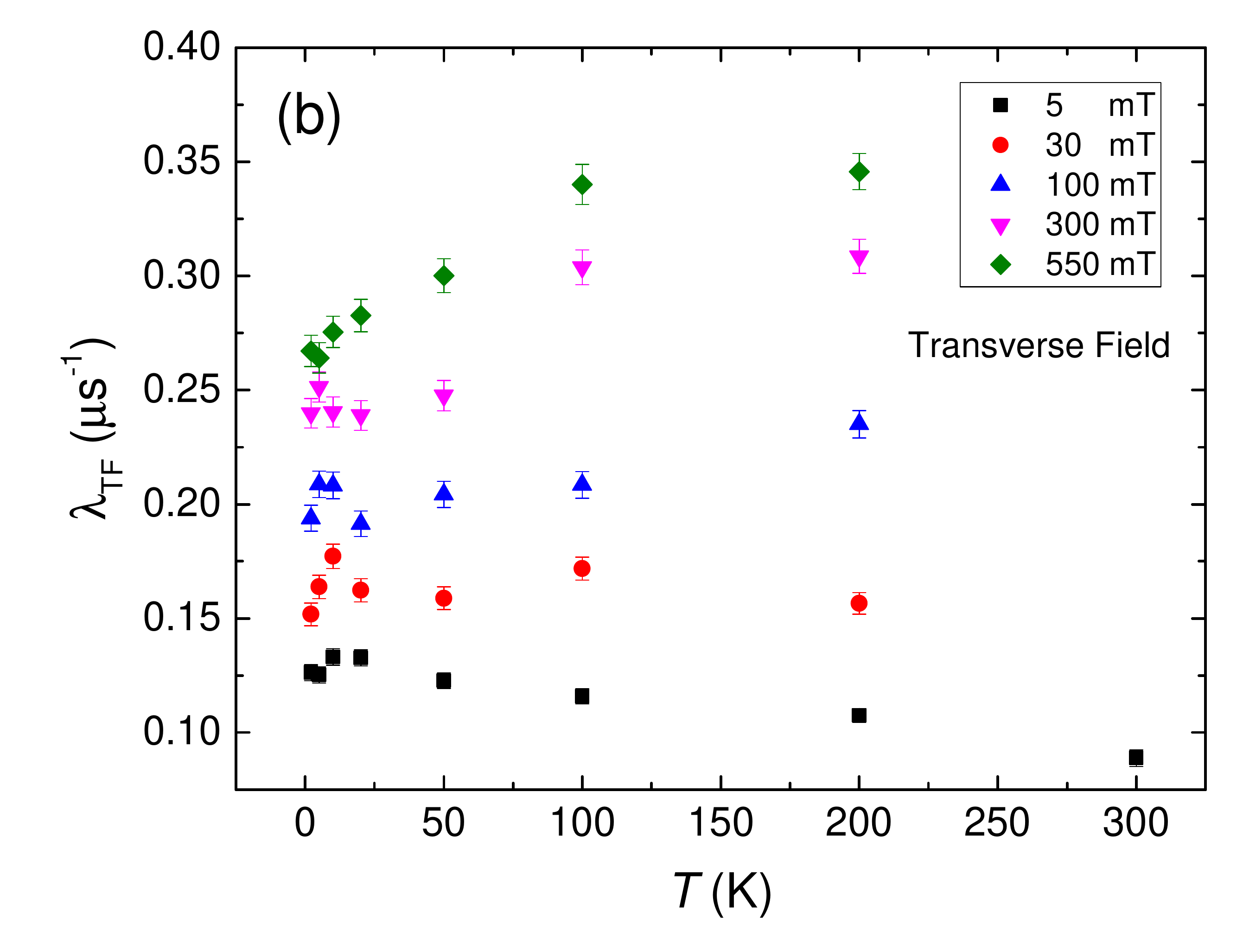}
	%\vspace{-7.5cm}
	\caption{\footnotesize (a) Temperature dependencies of ZF, and (b) temperature and field dependence of TF-$\mu$SR relaxation rates in Si$_{3}$N$_{4}$.}
	\label{Fig:Si3N4Field}
\end{figure}

In order to characterize Si$_3$N$_4$, the material available from the supplier \cite{Si3N4supplier} was studied in magnetization and $\mu$SR experiments. The magnetization measurements did not detect any presence of the ferromagnetic component. The results of the zero-field  and transverse-field $\mu$SR experiments of Si$_3$N$_4$ are summarized in Figure~\ref{Fig:Si3N4Field}. Experiments were performed for temperatures ranging from 1.6 to 300~K and fields from 0 up to 0.5~T at the General Purpose Surface decay instrument (GPS, PSI, Switzerland).

Figure~\ref{Fig:Si3N4Field}~(a) shows the temperature dependence of the ZF-$\mu$SR signal relaxation rates obtained from the fit of the following two-component function to the time dependent muon-spin polarization:
 \begin{eqnarray}
A_{0}P(t)&=& A_{\rm fast}\exp(-\lambda_{\rm fast} t) + A_{\rm slow}\exp(-\lambda_{\rm slow} t)~.
 \label{eq:ZF}
 \end{eqnarray}
Here $A_{0}$ is the initial asymmetry and $P(t)$ is the the muon-spin polarization. $A_{\rm fast}/A_{\rm slow}$ and $\lambda_{\rm fast}/\lambda_{\rm slow}$ are intimal asymmetries and exponential relaxations of the fast/slow component, respectively. The major contribution ($\gtrsim 90$\%) came from the slow-relaxing part which is characterized by an almost temperature independent relaxation rate with $\lambda_{\rm slow}\simeq 0.12$~$ \mu $s$^{-1}$.

The TF-$\mu$SR time-spectra were found to be well fitted with:
  \begin{eqnarray}
A_{0}P(t)&=& A_{\rm 0}\exp(-\lambda_{\rm TF} t) \cos(\gamma_{\mu} B_{\rm ap} t + \varphi)~.
 \label{eq:TF}
 \end{eqnarray}
The temperature dependence of $\lambda_{\rm TF}$ is rather weak and it increases with increasing field by reaching $\sim 0.3$~$\mu$s$^{-1}$ ay 0.55~T (see Figure~\ref{Fig:Si3N4Field}~(b)). Note that $\lambda_{\rm TF}$ stay in the range of 0.1-0.35 $\mu$s$^{-1}$ allowing to use pistons made of Si$_3$N$_4$ in low-background $\mu$SR under pressure experiments.

 \section{Scientific example: pressure induced superconductivity in binary helimagnet  CrAs} \label{sec:example}

Reaching high pressures by using large sample volume cells made of strong and nonmagnetic materials is an engineering challenge. The necessity of using a low background pressure cell in $\mu$SR measurements is caused by a large fraction of muons stopped in cell walls and contributing the background signal. The polarization function describing the background signal needs to have a simple dependence on the field and temperature.

\begin{figure}[h]
	\centering
	\includegraphics[width=0.49\linewidth]{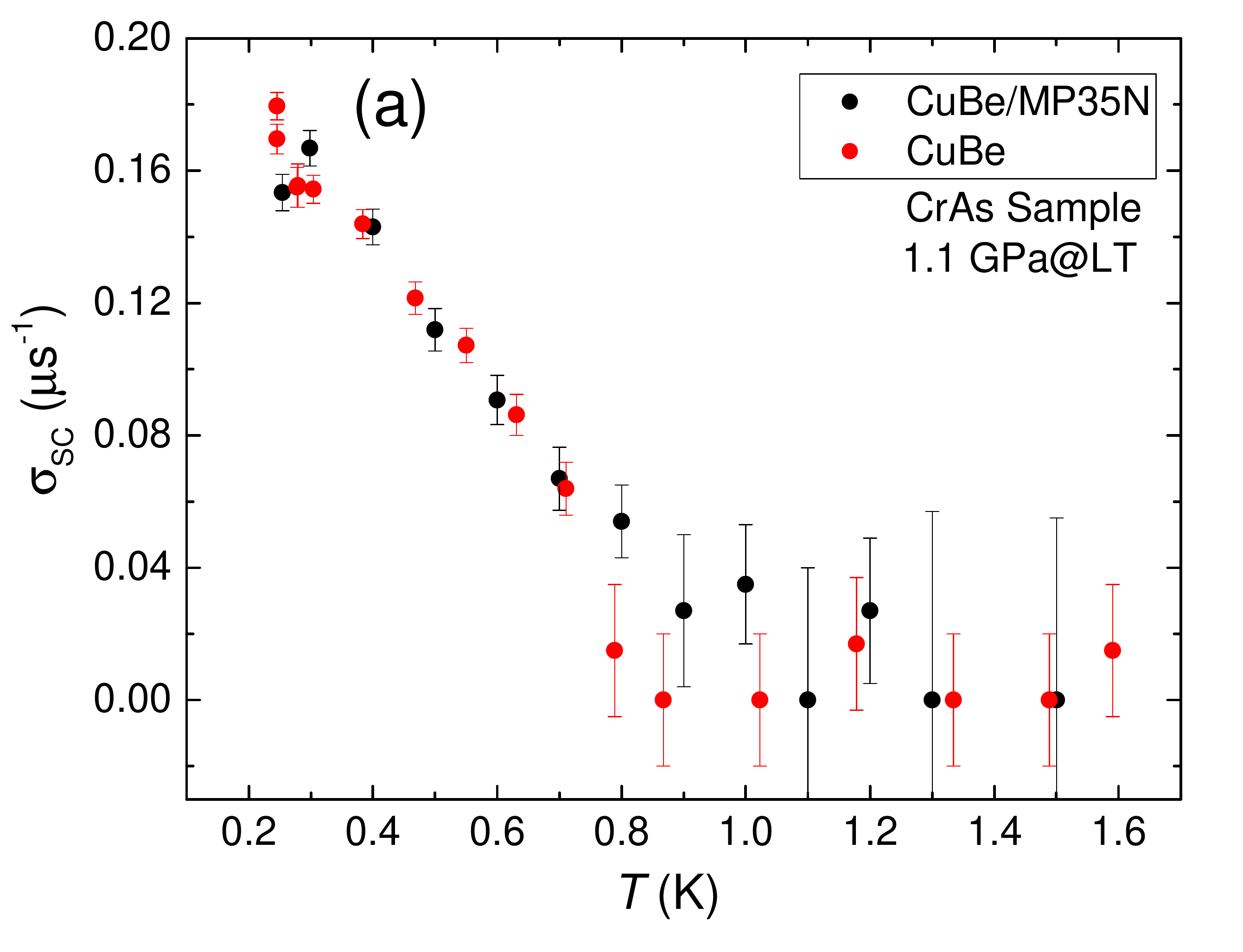}
	\includegraphics[width=0.49\linewidth]{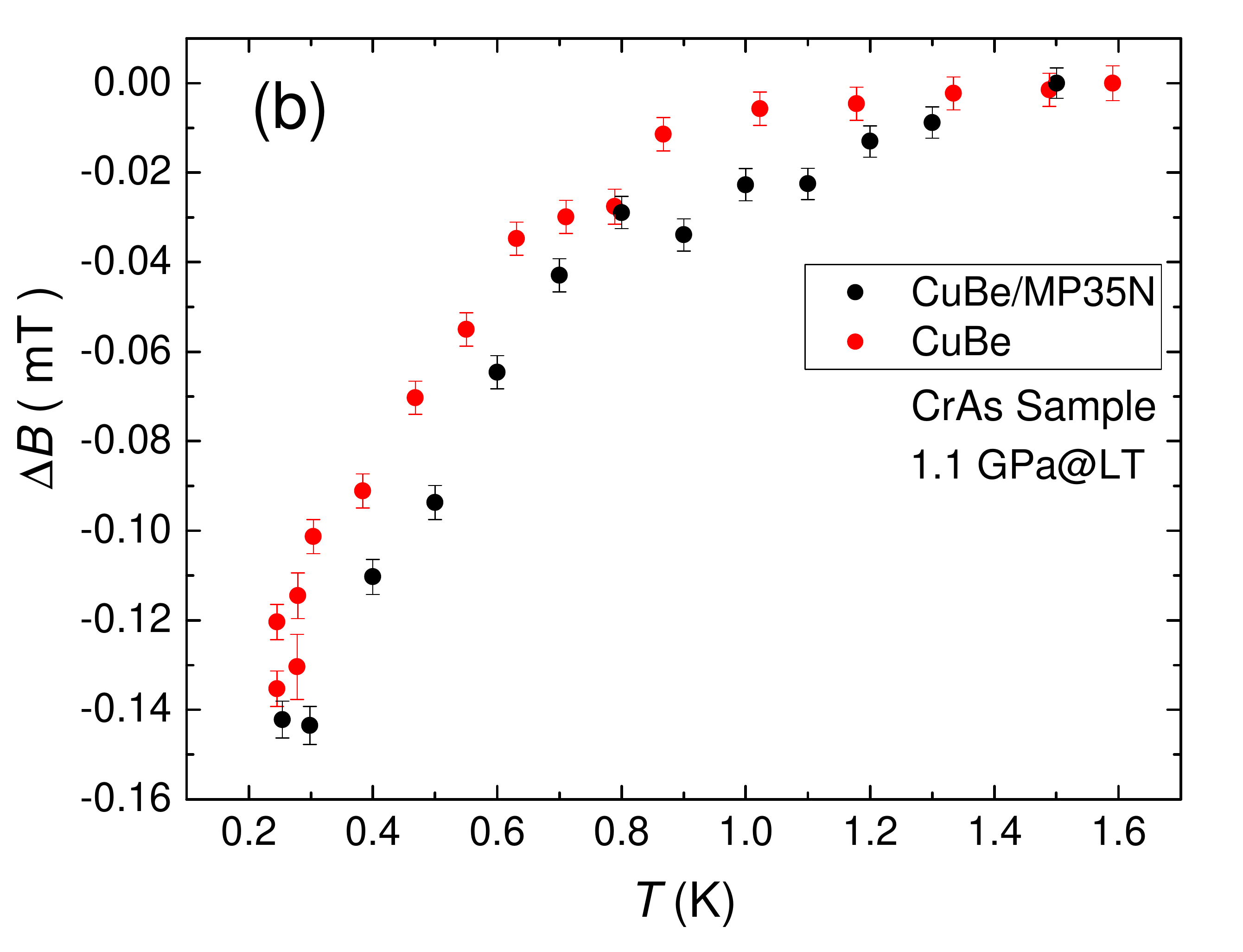}
	%\vspace{-7.5cm}
	\caption{\footnotesize Temperature dependencies of (a) Superconducting TF-$\mu$SR relaxation rates, and (b) field shift at $B_{ext}$~=~30~mT in CrAs sample using CuBe/MP35N double-wall cell and CuBe single-wall cell Ref.~\cite{KhasanovCrAs}.}
	\label{Fig:SigmaCompare}
\end{figure}

The test of the new CuBe/MP35N cell was performed by studying effect of pressure induced superconductivity in binary helimagnet CrAs~\cite{Wu}. The previous $\mu$SR experiments were performed up to $\simeq1.1$~GPa by using single-wall pressure cell made of purely CuBe alloy \cite{KhasanovCrAs}. In this compound the bulk magnetism is found to exist up to $p\simeq 0.35$~GPa.   Above 0.35~GPa the bulk magnetism was suppressed gradually and superconductivity with its maximum $T_{\rm c} \simeq 1.2$~K developed in parts of the sample remaining nonmagnetic down to low temperatures. Purely non-magnetic state is observed above 0.7 GPa.

In order to make comparison with the previously published data, the CrAs sample was placed inside the cell, the pressure $\simeq 1.1$~GPa was applied and TF-$\mu$SR experiments with the 30~mT applied field were performed. The obtained TF-$\mu$SR time spectra were analysed by using the polarization function:
  \begin{eqnarray}
A_{0}P(t)_{\rm TF}&=& A_{\rm s}e^{-(\sigma_{\rm SC}^{2}+\sigma_{\rm PM}^{2})t^{2}/2} \cos(\gamma_{\mu}B_{\rm s}t+\varphi)+A_{\rm bg}P(t)^{\rm bg}_{\rm TF}.
\label{eq:CrAsTF}
\end{eqnarray}
Here indices 's` and 'bg` denote the sample and the background contributions, respectively; $\sigma_{\rm SC}$ is the Gaussian relaxation rate caused by formation of the flux-line; and $\sigma_{PM}$ represents the nuclear moment contribution. $P(t)^{\rm bg}_{\rm TF}$ is the background polarization function expressed by Eq.~(\ref{eq:PcellBGSTF}). The Gaussian and exponential relaxation rates for background signal, as they reported in the previous section, were inserted into the $P(t)^{\rm bg}_{\rm TF}$ polarization function and left fixed for corresponding temperatures. The parameters related to the sample signal were left free during the fitting procedure.

The results of the fit together with the previously reported data are summarized in Figure~\ref{Fig:SigmaCompare}. $\sigma_{SC}$ increases below the superconducting transition temperature ($T_{\rm c}\simeq 0.9$~K) by reaching the value of $\simeq 0.16$~$\mu$s$^{-1}$ at $T\simeq 0.25$~K. The internal field inside the sample decreases below $T_{\rm c}$. Both these parameters behave as is expected for the type-II superconductor in the vortex state (see {\it e.g.} Ref.~\cite{Khasanov_05} and references therein).
More important, is that the parameters obtained by analysing CrAs $\mu$SR data for the sample placed inside the double-wall CuBe/MP35N or single-wall CuBe pressure cells remain almost the same. This demonstrates that weakly damped muSR spectra can be studied using the pressure cell reported here.

\section{Summary} \label{sec:conclusions}

To summarize, a low-background piston cylinder type hybrid pressure cell has been successfully designed, produced and commissioned. The cell is made out of CuBe and MP35N nonmagnetic alloys with the design and dimensions specifically adapted for muon spin rotation/relaxation measurements. The mechanical characteristics of the pressure cell were examined using the finite-element analysis. By including the measured stress-strain characteristics of the materials into the FEA model, the cell dimensions were
optimized with the aim of reaching the highest possible pressure while maintaining the
largest sample space (6mm diameter and 12mm length). The present design of the double-wall piston–cylinder pressure cell with an outer MP35N sleeve and inner CuBe cylinder allows one to reach pressures of up to $\simeq2.6$~GPa at room temperature ($\simeq 2.2$~GPa at low $T$) without any irreversible plastic deformation of the pressure cell. Additional $\mu$SR measurements were specifically dedicated to the Si$_{3}$N$_{4}$ ceramic material and the empty pressure cell to obtain the background signal necessary for the data analysis. The test $\mu$SR experiments were performed on CrAs compound. The experimental data were compared with those obtained earlier with the use of CuBe single-wall cell. The analysis of the old and new $\mu$SR data proved that muons stopped mainly in the sample and in the inner CuBe cylinder, resulting in a low-background $\mu$SR signal.

\vspace{0.5cm}

\noindent {\bf Acknowledgments}\\
 The authors acknowledges helpful discussions with S.~Klotz, E.~Leli\`{e}vre-Berna and B. Annigh\"{o}fer for helpful discussions and suggestions.

\vspace{0.5cm}
 \noindent {\bf Disclosure statement}\\
 No potential conflict of interest was reported by the authors.

\vspace{0.5cm}
 \noindent {\bf Funding}\\
  The work was performed at the Swiss Muon Source (S$\mu$S), Paul Scherrer Institute (PSI, Switzerland) within the framework of Horizon 2020 - INFRADEV Proposal No. 654000 'World class Science and Innovation with Neutrons in Europe 2020 (SINE2020)'.

\vspace{0.5cm}
 \noindent {\bf Additional Information}\\
 Correspondence and requests for materials
should be addressed to zurab.shermadini@psi.ch, rustem.khasanov@psi.ch, or matthias.elender@psi.ch

\end{document}